\documentclass[12pt,iop,twocolumn,aps,letter]{emulateapj}
\usepackage{amsmath}
\usepackage{amsfonts}
\usepackage{amssymb}
\usepackage{graphicx}
\usepackage{graphics}
\usepackage{threeparttable}
\usepackage{color}
\usepackage{bm}
\usepackage{subfigure}

\usepackage[pdfauthor={Wing-Kit Lee},%
pdftitle={FEATHERING INSTABILITY OF SPIRAL ARMS. II: PARAMETER STUDY},%
pdfsubject={FEATHERING INSTABILITY OF SPIRAL ARMS},
pagebackref=false,%
colorlinks,
citecolor=blue,
linkcolor=blue,
urlcolor=blue,
breaklinks,
]{hyperref}

\shorttitle{PARAMETER STUDY OF FEATHERING INSTABILITY}
\shortauthors{LEE}

\newcommand{\uhat}{\hat{u}}

\newcommand{\shat}{\hat{\sigma}}

\begin{document}
\title{FEATHERING INSTABILITY OF SPIRAL ARMS. \\ II. PARAMETER STUDY}
\author{Wing-Kit Lee$^{1,2}$}
\affil{$^1$Center for Astrophysics and Space Sciences \\
University of California, San Diego, La Jolla, CA 92093-0424\\}
\affil{$^2$Institute of Astronomy and Astrophysics, Academia Sinica, Taipei 115, Taiwan, Republic of China}
\email{wklee@asiaa.sinica.edu.tw}

\defcitealias{LS2012}{Paper I}	
\defcitealias{SMR1973}{SMR}
\defcitealias{1986ApJ...309..496L}{LCB}

\begin{abstract}
We report the results of a parameter study of the feathering stability in the galactic spiral arms. A two-dimensional razor-thin magnetized self-gravitating gas disk with an imposed two-armed stellar spiral structure is considered. Using the formulation developed previously by Lee and Shu, a linear stability analysis of the spiral shock is performed in a localized Cartesian geometry. Results of the parameter study of the base state with a spiral shock are also presented. The single-mode feathering instability that leads to growing perturbations may explain the feathering phenomenon found in nearby spiral galaxies. The self-gravity of the gas, characterized by its average surface density, is an important parameter which 1) shifts the spiral shock further downstream; 2) increases the growth rate and decreases the characteristic spacing of the feathering structure due to the instability. On the other hand, while the magnetic field suppresses the velocity fluctuation associated with the feathers, it does not strongly affect their growth rate. Using a set of typical parameters of the grand-design spiral galaxy M51 at 2 kpc from the center, the spacing of the feathers with the maximum growth rate is found to be 530 pc, which agrees with the previous observational studies.
\end{abstract}

\section{Introduction}
Feathers are commonly found as dust extinction feature in the nearby spiral galaxies \citep[e.g.,][]{1970IAUS...38...26L,2006ApJ...650..818L}. They jut out almost perpendicularly from the major spiral arms to the interarm region. The feathers also differ from the stellar ``spurs" in galaxies \citep{1980ApJ...242..528E} because they are traceable only in the regions with less active star formation (e.g., not obscured by giant H\,II regions), and become undetectable farther away from the spiral arm \citep{2006ApJ...650..818L}. Optical images of M\,51 \citep[e.g.,][]{2001AJ....122.3017S} showed the feathering phenomenon is more distinguishable in the inner region of the galaxy than the outer part where star-forming regions are prominent (see, Figure \ref{fig:optical}).

The archival study by \citet{2006ApJ...650..818L} identified a few characteristics of the feathers, namely, their spacing is larger at the larger radius with lower molecular gas surface density; and they are commonly associated with dense molecular gas inferred from CO observations \citep[see also,][]{2008ApJ...689..148C}. This suggests the important role of gravitational instability and is consistent with the suggestion that the feather may share the same origin as stellar spurs \citep{1980ApJ...242..528E}. Yet, little is known about whether the feathers are regular and periodic in nature nor if there is any relation between the feathers and the star formation along the spiral arm. Young star complexes are found embedded in the dark dust filaments in the spiral arms \citep{2014ApJ...780...32E}, which supports the relation between feathers and the formation of molecular clouds and stars in the early stage.

Theoretical studies of spiral arm substructure have been primarily based on the shearing instability, which is due to both the background shear of the galactic differential rotation and the shear due to the stellar spiral gravitational potential. The gas shock resulted in the latter case provides a post-shock shear in a reverse sense compared to the galactic shear. \citet{KO2002} performed two-dimensional shearing-box magneto-hydrodynamic (MHD) simulations and found a good agreement between their streamlines and the calculation of a modified shearing analysis with magnetic field based on \citet{B88}. The gas response to the spiral structure in the two-dimensional purely-hydrodynamic simulations of the whole galactic disk is unstable to the shearing instability if the spiral forcing is too strong \citep{2004MNRAS.349..270W, 2006ApJ...647..997S, 2014MNRAS.440..208K} or if the gas is too cold \citep{2006MNRAS.367..873D}. \citet{2006ApJ...647..997S} found the gas response remains stable in the presence of a magnetic field, but another MHD instability arises when the gas self-gravity is important. As the interstellar medium (ISM) in the galactic scale is magnetized, the self-gravitating MHD instability found in numerical simulations may correspond to the feathering phenomenon in real galaxies.

With this line of thought, \citet[][hereafter Paper I]{LS2012} formulated the formation of substructure of the galactic spiral arms by considering an intrinsic instability in the spiral shock, and provided a connection between this feathering instability and self-gravitating magnetized ISM in the galactic scale. The spiral shock occurs when the gas is passing through the stellar spiral potential \citep{Roberts1969a}, where the spiral structure itself is induced by the spiral density wave \citep{1964ApJ...140..646L}. The spiral forcing is not necessarily very strong compared centripetal force, which was estimated to be $F=5\%$ for the Milky Way \citep{1969ApJ...158..889Y}, where $F$ is the ratio between spiral and centripetal accelerations. The gas response to the underlying two-armed spiral structure (TASS) is nonlinear and has been studied numerically \citep[e.g.,][hereafter SMR]{Roberts1969a, SMR1973}. This time-steady, quasi-one-dimensional TASS state with a spiral shock is the base state of the instability in our study.

Regarding the origin or longevity of the spiral density waves, there are two competing views proposed. The first is the hypothesis of quasi-steady spiral structure (QSSS) \citep{1964ApJ...140..646L, 1996ssgd.book.....B}, in which the spiral pattern is long lived and is a normal mode of the thin stellar disk. The second is the superposition and nonlinear saturation of the growing modes \citep{2012ApJ...751...44S, 2014ApJ...785..137S} that give rise to the transient but recurrent spiral patterns. The latter scenario is based on N-body simulations of a cold, unbarred, collision-less stellar disk. Similar study by \citet{2013ApJ...766...34D} showed that the spiral patterns are not material entities but statistically long-lived density waves induced by local density perturbations \citep[c.f.,][]{1966ApJ...146..810J}. In particular, it is very difficult to produce persistent grand-design two-arm spiral structure in these simulations of an isolated stellar disk \citep{2011MNRAS.410.1637S}. An early review by \citet{1995NYASA.773..125L} remains relevant in explaining the importance among different approaches. As in \citetalias{LS2012}, we do not address different generation mechanisms of the base state of spiral structure, but focus on the formation of spiral substructure as a response to the steady forcing associated with the classic QSSS picture. Therefore, our findings of the instability in the self-gravitating, magnetized ISM should not change significantly if the stellar spiral pattern is only statistically persistent.

We demonstrated in \citetalias{LS2012} the existence of the unstable mode of instability that results in the feather-like density structure in the post-shock region. Such feathering instability is driven by the gas self-gravity in the spiral shock and complemented by the expanding set of magnetic fields in the interarm region. Very recently, \citet{2014ApJ...789...68K} performed a stability analysis for the purely-hydrodynamic case without magnetic field and self-gravity, and reported that the wiggle instability \citep[c.f.,][]{2004MNRAS.349..270W} is unlikely to produce feathers that are a few hundreds pc apart as observed. In any case, normal-mode analysis such as \citetalias{LS2012} and \citet{2014ApJ...789...68K} explored different physical regimes in the system and provided the length scale of the corresponding instabilities. Therefore, this parameter study is crucial for understanding the feathering instability and to provide better diagnostics of future numerical simulations and observations. 

The paper is organized as follows. In Section \ref{SecBE}, we review the basic equations. Our set-up is essentially the same as \citetalias{LS2012}. In Section \ref{SecPhyScale}, we discuss the free parameters and summarize their meaning in Table \ref{c3table1}. As the TASS base state and the feathering perturbation are coupled, we first study the parameter dependence of the TASS base state in Section \ref{SecBasic}. In Section \ref{SecFeather}, we give the results for the feathering instability. We summarize and discuss our findings in Section \ref{SecDiscuss}. Lastly, we have the conclusion in Section \ref{Sec:Conclusion}.

\section{Basic Equations}\label{SecBE}

The MHD response of a self-gravitating razor-thin gas disk under the influence of a stellar two-arm spiral pattern is studied. We start from the ideal MHD equations in a rotating frame of reference. This frame rotates at the pattern speed of the stellar spiral structure, such that its gravitational forcing is static. We transform the system from the usual cylindrical coordinates ($\varpi$, $\varphi$, $z$) to local Cartesian coordinates ($\eta$, $\xi$, $z$). Such coordinate transformation was introduced by \citet{Roberts1969a} for his calculation concerning the quasi-one-dimensional spiral shock problem along the perpendicular direction of the spiral arm (i.e., the TASS state). The framework was extended to study magnetic field \citep{1970ApJ...161..887R}, two-phase ISM \citep{1972ApJ...173..557S}, spiral forcing and ultra-harmonic resonances \citepalias{SMR1973}, gas self-gravity \citep[][hereafter LCB]{1986ApJ...309..496L}, and more recently to estimate the corotation radius \citep{2004MNRAS.349..909G}. In \citetalias{LS2012}, we included both gas self-gravity and magnetic field in the calculation which were not considered simultaneously in previous analytical treatments.

The local quasi-one-dimensional TASS calculation is naturally extended into two-dimensions by considering the direction parallel to the stellar spiral arm as well. We adopt the tight-winding approximation of the spiral structure (or equivalently a WKB approximation) which allows a simple linear form of the gravitational potential and forcing of the stellar density wave \citep{1964ApJ...140..646L}. Under this asymptotic approximation (i.e., $\sin i$ is small, where $i$ is the pitch angle of the spiral arm), the calculation domain is simplified into a local rectangular box with the one side aligned parallel with the spiral arm and periodic in both parallel and perpendicular directions (see Figure \ref{fig:coord}). We also ignore the galactic shear within the box (unlike other local analysis using shearing coordinates, e.g., \citet{B88,KO2002}), but retain the Coriolis terms. This simplification is necessary in order to remove the explicit time dependence of the quasi-radial boundaries and provide the periodic boundary conditions suitable for the normal-mode analysis. While non-self-gravitating and purely-hydrodynamic instability reported in some simulations may be relevant in certain situation and contribute to interstellar turbulence, we focus on the formation of spiral arm substructure through an asymptotic treatment to study the instability induced by self-gravity and modified by interstellar magnetic fields. Therefore, the comparison between the feathering instability and other shearing instability will be left for future investigation. 

For completeness, we briefly review the basic equations that were previously derived in \citetalias{LS2012}. Interested readers should consult \citetalias{LS2012} for the details of the following: 1) non-dimensionalization of the basic ideal MHD equations; 2) the expression of Lorentz force in the TASS state calculation; 3) the boundary conditions or equivalent shock-jump conditions; and 4) linearization of equations for stability analysis. Readers who are only concerned with the results of this parameter study may skip ahead to the next section.

\subsection{Dimensional Equations in Local Coordinates}

\begin{figure}[!htb]
\begin{center}
\subfigure[Optical image of M\,51 from the Hubble Space Telescope]{
\makebox[.2\textwidth]{
\includegraphics[scale=1.0]{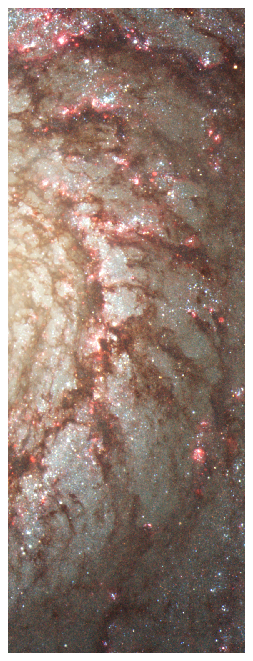}
\label{fig:optical}
}
}
\subfigure[Local coordinate system]{
\makebox[0.2\textwidth]{
\includegraphics[scale=0.45]{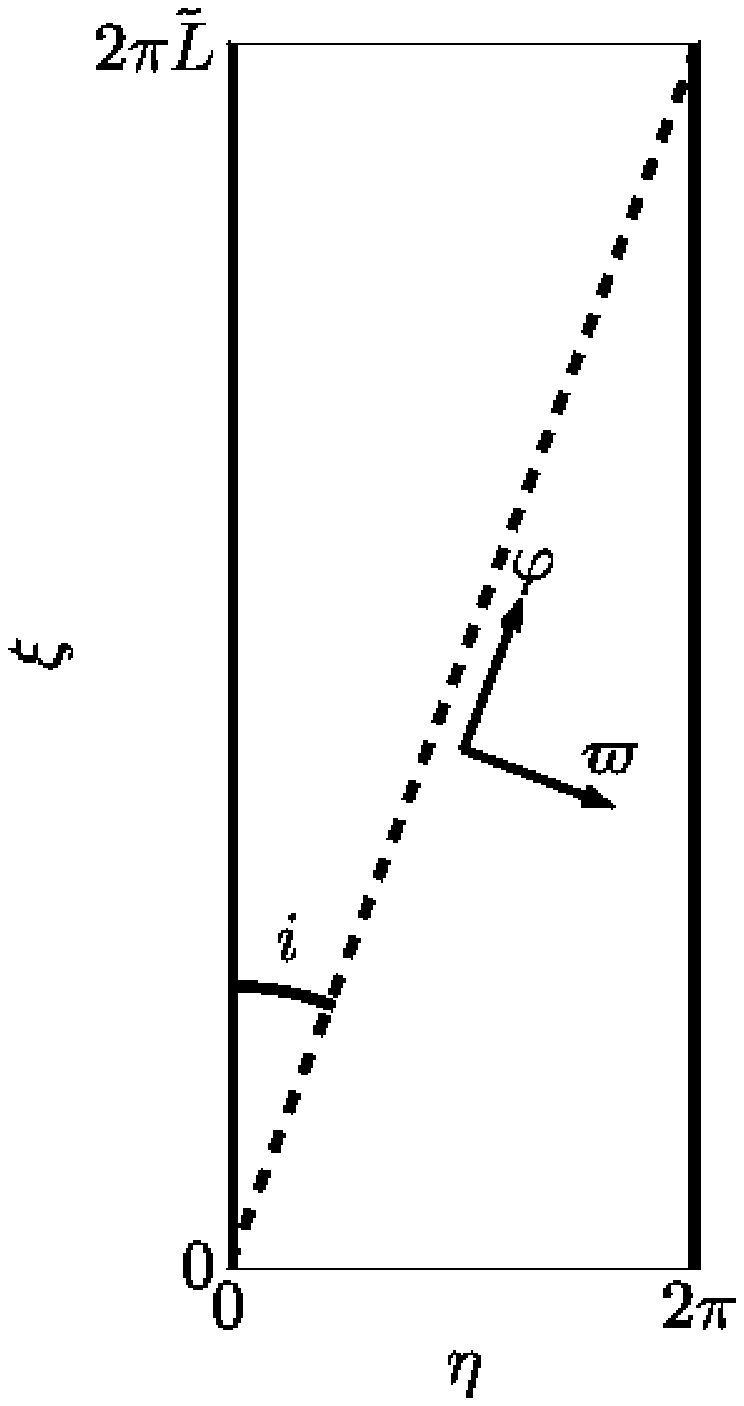} %c3fig1.eps %fig5b.eps
\label{fig:coord}
}
}
\caption{(a) We show a small segment of a spiral arm with some dark dusty feathers jutting out from the primary dust lane on the left. (b) The horizontal and vertical directions correspond to the perpendicular and parallel directions to the spiral arm, respectively. The spiral arms are located at the $\xi$-axis at both $\eta=0$ and $2\pi$. The diagonal dashed line indicates the azimuthal direction.}
\end{center}
\end{figure}

A local, doubly-periodic rectangular box is constructed to align with the spiral arm. The axes are defined by the local Cartesian coordinates ($\eta$, $\xi$) used by \citetalias{SMR1973}, which is basically rotationally-transformed from the usual cylindrical coordinate system ($\varpi$, $\varphi$). To be clear, $\varpi$ and $\varphi$ are the radial and azimuthal coordinates in a rotating frame centered at the galaxy center, respectively. The $\eta$-coordinate goes from 0 to $2\pi$, which corresponds to the perpendicular displacement from one spiral arm to the next. The $\xi$-coordinate has the same scale as $\eta$ but runs in the parallel direction to the spiral arm. Since the arm-to-arm distance is $L_{\rm arm} = 2\pi \varpi \sin{i}/m$, where $m$ is the number of spiral arms, the physical length scale of $\eta$- and $\xi$-coordinates is
\begin{align}
L_0=\varpi \sin{i}/m.
\end{align}
Therefore, the two coordinate systems are related by the following metric:
\begin{align}
ds^2 = d\varpi^2 + \varpi^2 d\varphi^2 = L_0^2 (d\eta^2 + d\xi^2),
\end{align}
such that the unit vectors are related by
\begin{align}
\hat{e}_\varpi &= \cos i\,\hat{e}_\eta - \sin i\, \hat{e}_\xi, \\ 
\hat{e}_\varphi &= \sin i\,\hat{e}_\eta + \cos i\, \hat{e}_\xi, 
\end{align}
where $\hat{e}$ is the unit vector pointing to each axis. Since $\eta$ and $\xi$ are Cartesian coordinates, the partial derivatives are simply their dimensional counterpart with a simple scaling. For example, the two-dimensional divergence of the velocity $\mathbf{u}$ is written as
\begin{align}
\nabla \cdot \mathbf{u} = L_0^{-1}\left(\frac{\partial u_\eta}{\partial \eta} + \frac{\partial u_\xi}{\partial \xi}\right),
\end{align}
where the curvature terms are dropped, and $u_\eta$ and $u_\xi$ are the $\eta$- and $\xi$-components of $\mathbf{u}$, respectively. On the other hand, the aspect ratio of the box is given by $\tilde{L} = \cot{i}$ such that the periodicity of $\xi$ is $2\pi\tilde{L}$. The ($\eta$, $\xi$) coordinate system is shown in Figure \ref{fig:coord}, where we ignore the curvature. 

In a rotating frame with angular frequency, $\Omega_{\rm p}$, the dimensional equations of continuity and conservation of momentum for ideal MHD read (c.f., Equations (33)-(35) in \citetalias{LS2012})
\begin{align}
\label{eq:dyn1}
\frac{\partial \Sigma}{\partial t} + \nabla \cdot(\Sigma \mathbf{u}) &=0, \\
\label{eq:dyn2}
\frac{\partial u_{\eta1}}{\partial t} + \mathbf{u} \cdot \nabla u_{\eta1} &= 2\Omega u_{\xi1} - \frac{1}{\Sigma L_0}\frac{\partial \Pi}{\partial \eta} - \frac{1}{L_0}\frac{\partial \mathcal{U}}{\partial \eta} + f_{\eta},\\
\label{eq:dyn3}
\frac{\partial u_{\xi1}}{\partial t} + \mathbf{u} \cdot \nabla u_{\xi1} &=-\frac{\kappa^2}{2\Omega} u_{\xi1} - \frac{1}{\Sigma L_0}\frac{\partial \Pi}{\partial \xi} - \frac{1}{L_0}\frac{\partial \mathcal{U}}{\partial \xi} + f_{\xi},
\end{align}
where $\Sigma$ is the gas surface density; $\mathbf{u}_1 =(u_{\eta1}, u_{\xi1})$ is the non-circular component of the velocity; $\Omega$ and $\kappa$ are the frequency for the circular rotation and the epicyclic motion, respectively; $\Pi$ is the vertically-integrated gas pressure; and $\mathbf{f}=(f_\eta, f_\xi)$ is the Lorentz force per unit mass. The total velocity in the pattern frame is given by
\begin{align}
\mathbf{u} = \mathbf{u}_0 + \mathbf{u}_1,
\end{align}
where $\mathbf{u}_0 = \varpi(\Omega(\varpi)-\Omega_{\rm p})\hat{e}_\varphi$ is the circular velocity. The circular velocity $\mathbf{u}_0(\varpi)$ is the solution to the axisymmetric state, which is in the hydrostatic equilibrium in the radial direction. Therefore, the right-hand-side of the momentum equations only contains the non-axisymmetric contribution. In particular, the total effective gravitational potential can be written as
\begin{align}
\mathcal{V}_{\rm total} = \mathcal{V}_0(\varpi) - \tfrac{1}{2}\varpi^2 \Omega_{\rm p}^2 + \mathcal{U}(\varpi, \varphi, z),
\end{align}
where contributions from axisymmetric potential $\mathcal{V}_0$ (i.e., due to bulge and dark halo, etc) and centrifugal force are cancelled out in Equations (\ref{eq:dyn2}) and (\ref{eq:dyn3}). The remaining non-axisymmetric gravitational potential $\mathcal{U}$ in the usual cylindrical coordinates is given by
\begin{align}
\mathcal{U} = \mathcal{V}_{\rm spiral}(\varpi,\varphi) + \mathcal{V}_{\rm gas}(\varpi,\varphi,z),
\end{align}
where $\mathcal{V}_{\rm spiral}$ is the potential of the stellar spiral structure, $\mathcal{V}_{\rm gas}$ is the self-gravity potential of the gas. The self-gravity of the gas is governed by the Poisson equation in a thin-disk geometry, which is given by
\begin{align}
\nabla ^2 \mathcal{V}_{\rm gas} = 4\pi G\Sigma(\varpi,\varphi)\delta(z),
\end{align}
where $G$ is the gravitational constant and $\delta$ is a Dirac-Delta function. Note that $z$-coordinate is generally suppressed except it is required for the Poisson equation for a thin-disk. The Lorentz force per unit mass in the ideal MHD limit at the mid-plane ($z=0$) is given by
\begin{align}\label{eq:lorentzforce}
\mathbf{f} = -\frac{z_0}{2\pi\Sigma}(\nabla \times \mathbf{B}) \times \mathbf{B},
\end{align}
where $\mathbf{B}$ is the magnetic field and $z_0$ is the scale-height of the gas disk. The time evolution of the magnetic field is governed by the induction equation for the ideal MHD. In practice, for the perturbational magnetic field $\mathbf{B}_1$, we solve the $z$-component of the magnetic vector potential $A_z=A_z(\eta,\xi)$ in lieu of the two-dimensional magnetic field \citepalias[c.f., Appendix B of][]{LS2012}, where $\mathbf{B}_1=\nabla \times (A_z \hat{e}_z)$. To close the problem, we adopt a logatroptic equation of state (EOS) \citep{1989ApJ...342..834L} for the turbulent gas, such that $\Pi_{\rm turb} = \Sigma_0 {\rm v}_{\rm t0}^2 \ln (\Sigma/\Sigma_0)$, where ${\rm v}_{\rm t0}$ is the mean velocity dispersion and $\Sigma_0$ is the average gas surface density, to mimic the lower velocity dispersion for denser ISM \citep[e.g.,][]{2005ApJ...629..849P}. 

The non-axisymmetric contributions in the momentum equations from the turbulent pressure and magnetic field are obtained by applying the tight-winding approximation. Consistent with this approximation where the radial wavenumber is large compared to its azimuthal counterpart, the radial variation of axisymmetric state such as $\Omega(\varpi)$ and $\kappa(\varpi)$ is ignored in the dynamical equations in local ($\eta$, $\xi$) coordinates (i.e., constant $\Omega(\varpi)$ across the domain). \citetalias{LS2012} provides a great detail of the treatment and we shall not repeat the formulation here.

\subsection{Non-dimensionalization}\label{SecNormalization}
To proceed, we introduce several normalization scales, which we use for the dimensionless parameters. We follow the formulation in \citet{Roberts1969a} and \citetalias{SMR1973} to introduce two velocity scales $U$ and $V$ for the normalization. We define
\begin{align}
U \equiv \frac{\varpi \Omega \sin{i}}{m} \quad \text{and}
 \quad V \equiv \frac{\varpi \kappa^2 \sin{i}}{2\Omega m}.
\end{align}
Thus, the normalization factors for the velocity in $\eta$- and $\xi$-directions are given by
\begin{align}
\sqrt{2UV} = \kappa L_0 \quad \text{and}
 \quad V = \frac{\kappa^2}{2\Omega}L_0.
\end{align}
Since the $\eta$-component of the background circular velocity is $u_{\eta0} = \varpi (\Omega - \Omega_{\rm p}) \sin{i}$, its dimensionless counterpart is $-\nu \equiv u_{\eta 0} / \sqrt{2UV} = m(\Omega - \Omega_{\rm p}) / \kappa$, which is the ratio of the frequency of Doppler-shifted circular flow to that of the epicyclic oscillations. The surface density is simply normalized by the mean surface density $\Sigma_0$.

We transform the $\eta$- and $\xi$-momentum equations into dimensionless ones by dividing them by $2UVL_0$ and $\sqrt{2UV}VL_0$, respectively. Consistent with the above normalization, the dimensionless time variable is defined by $d\tau = \kappa dt$. In addition, the self-gravity potential of the gas can be rewritten as $\mathcal{V}_{\rm gas} = 2\pi G\Sigma_0 L_0 \phi$, where $\phi$ is the dimensionless counterpart of $\mathcal{V}_{\rm gas}$. For the gravitational force due to the stellar spiral, its dimensionless counterpart is 
\begin{align}
- \frac{1}{L_0}\frac{\partial \mathcal{V}_{\rm spiral}}{\partial \eta} \rightarrow - f\sin \eta,
\end{align}
where $f$ is the spiral forcing parameter defined in Section \ref{SecPhyScale} and we adopt a simple sinusoidal form for the linear stellar spiral density wave where the minimum of gravitational potential is located at $\eta=0$. 

\subsection{Summary of Basic Equations}

In summary, Equations (\ref{eq:dyn1}) through (\ref{eq:lorentzforce}) along with an induction equation for ideal MHD govern the gas response to an imposed stellar spiral structure. The axisymmetric equilibrium state is the radial hydrostatic equilibrium, in which the centripetal acceleration $\varpi \Omega^2$ is balanced by the radial gravity due to axisymmetric contribution of the gravitational potential $V_0(\varpi)$ and pressure gradient due to turbulent and magnetized gas. Therefore, adaptation of a particular rotation curve of the galaxy (i.e., $\varpi\Omega(\varpi)$) automatically assumes such equilibrium for the purely circular flow. For the time-steady, quasi-one-dimensional calculation of the nonlinear TASS state, the time and $\xi$-derivatives are dropped. This reduces to the same set of governing equations of nonlinear TASS state studied in \citetalias{SMR1973} except for the different expressions of the effective sound speed and force terms for magnetic field and gas self-gravity. As the spiral structure is non-axisymmetric, this will induce the variation of radial velocity (and hence the $\eta$ component) and lead to a spiral shock. We shall discuss the dimensionless parameters in the next section and continue the discussion of the feathering instability in Section \ref{SecFeather}.

\begin{table}[!htb]
\small
\begin{center}
\caption[Description of Parameters]{Description of Parameters}\label{c3table1}
Group I \\
\begin{tabular}{cc}
\hline
$\nu$ & background gas velocity perp. to the arm \\
$f$ & strength of the stellar spiral potential \\
$\alpha$ & strength of self-gravity of the gas \\
$x_{\rm t0}$ & square of turbulent speed of the gas \\
$x_{\rm A0}$ & square of Alfv\'{e}n's speed \\
\hline
\end{tabular}
\\[1.5ex]
Group II \\
\begin{tabular}{cc}
\hline
$\Omega/\kappa$ & ratio of rotational and epicyclic freq.\\
$\tan{i}$ & tangent of pitch angle of the spiral arm \\
\hline
\end{tabular}
\\[1.5ex]
Group III \\
\begin{tabular}{cc}
\hline
$L_{\rm arm}$ & perpendicular separation of spiral arms \\
$z_0$ & half-height of the gas disk \\
$\Omega_{\rm p}$ & pattern speed \\
\hline
\end{tabular}

\end{center}
\end{table}

\section{Parameters and Physical Scales}\label{SecPhyScale}
There are 7 dimensionless parameters in the local analysis, namely: the background gas velocity perpendicular to the arm, $\nu$; the strength of stellar spiral potential, $f$; the strength of self-gravity of the gas, $\alpha$; the square of turbulent speed of the gas, $x_{\rm t0}$; the square of Alfv\'en's speed, $x_{\rm A0}$; the ratio of the rotational and epicyclic frequency, $\Omega/\kappa$; and the tangent of pitch angle of the spiral arm, $\tan{i}$. To investigate the effects of each parameter effectively, the parameters are separated into 3 groups according to their roles: Group I is a set of dimensionless parameters that determines the TASS profile; Group II is the set of dimensionless parameters that sets the problem of the feathering perturbation, in addition to a given set of Group I parameters; Group III is a set of dimensional scales that sets the physical scales and units. Thus, the dimensional parameters are defined separately such that the dimensionless calculations can be scaled to match different physical conditions. We adopt the same notation of variables as \citetalias{LS2012}. Same as previous studies, we are primarily concerned with the region inside the corotation radius such that there is a stronger spiral shock \citepalias{SMR1973}, where $\Omega > \Omega_{\rm p}$ and $\nu < 0$. 

The conversion formulas between some physical variables and the aforementioned dimensionless parameters are presented below. In particular, the perpendicular distance between spiral arms, $L_{\rm arm}$ and the pattern speed, $\Omega_{\rm p}$ are used to obtain the dimensional scales of length and time, respectively. Except for the background magnetic field, $B_{\varphi 0}$ (which also depends on the half-height of gas disk, $z_0$), most of the variables can be scaled with a physical unit accordingly using $L_{\rm arm}$ and $\Omega_{\rm p}$.
\paragraph*{Group I} The dimensionless parameters in Group I specify the TASS state. They are defined by the following:
\begin{align}\label{eq5}
\nu &\equiv m (\Omega_{\rm p} - \Omega)/\kappa, \\\label{eq6}
f &\equiv \left(\frac{\Omega}{\kappa}\right)^2 \left(\frac{m F}{\sin{i}}\right), \\\label{eq7}
\alpha &\equiv \frac{2 \pi m G\Sigma_0}{\varpi \kappa^2 \sin{i}}, \\\label{eq8}
x_{\rm t0} &\equiv \frac{{\rm v}^2_{\rm t0}}{2UV}, \\\label{eq9}
x_{\rm A0} &\equiv \frac{{\rm v}^2_{\rm A0}}{2UV},
\end{align}
where ${\rm v}_{\rm t0}$ and ${\rm v}_{\rm A0}$ are the dimensional turbulent speed of the gas and the Alfv\'{e}n's speed, respectively, and $\sqrt{2UV} = \varpi \kappa \sin{i}/m$ is the normalization factor for velocities in the perpendicular direction to the spiral arm, and $F$ is the ratio between the stellar spiral forcing and centripetal force \citepalias[c.f., Equation 8 of][]{LS2012}. We set the value $f$ large enough such that a spiral shock exists \citepalias[c.f.,][]{SMR1973}.

\paragraph*{Group II} The dimensionless parameters in Group II are $\tan{i}$ and $\Omega/\kappa$. In general, these two parameters are not completely arbitrary in a sense that we usually have good measurements of the pitch angle of a spiral arm and the rotation curve.

\paragraph*{Group III and Other Dimensional Variables} The parameters in this group are physical length and time scales. The perpendicular separation between spiral arms, $L_{\rm arm}$, the pattern speed, $\Omega_{\rm p}$ and the half-height of the gas disk, $z_0$ are used. Equivalently, other dimensional parameters can be specified, such as galacto-centric radius $\varpi$, and rotational frequency $\Omega(\varpi)$ (or, $\kappa(\varpi)$) for the purpose of dimensional conversion. Thus, the dimensional scales of the gas surface density and magnetic field are also set. From Equation (\ref{eq7}), the mean gas surface density can be written as $\Sigma_0 = \alpha\Sigma_{\rm A}$, where
\begin{align}
\Sigma_{\rm A} \equiv \left(\frac{\varpi\sin{i}}{m}\right)\frac{\kappa^2}{2\pi G} = \frac{\kappa^2 L_{\rm arm}}{4\pi^2 G}
\end{align}
is a scale of gas surface density set by the galactic parameters. In general, $\Sigma_{\rm A}$ is large compared to the realistic gas surface density. Using the typical numbers for the inner part ($\varpi=2\,{\rm kpc}$) of M$\,$51, we have $\Sigma_{\rm A} = 460\,{\rm M_\odot\,pc^{-2}}$. Thus, we expect the value of $\alpha$ is in the order of 0.1. For the background (circular) magnetic field, we have 
\begin{align}\label{eq:defB0}
B_{\varphi 0} &= (\alpha x_{\rm A0})^{1/2}\left(\frac{\varpi\sin{i}}{m}\right)^{3/2}\frac{\kappa^2}{(Gz_0)^{1/2}},
\end{align}
where $z_0$ is the scale-height of the gas disk. Using the numbers for M$\,$51, we have
\begin{align}
B_{\varphi0} = 210\,(\alpha x_{\rm A0})^{1/2} \mu{\rm G},
\end{align}
where we take $z_0 = 200\,{\rm pc}$.
For comparison to previous numerical simulations, such as \citet{KO2002,2006ApJ...646..213K}, we provide the conversion formulae for the Toomre's parameter $Q_0$ and plasma beta $\beta_0$:
\begin{align}\label{eq:defQ0}
Q_0 &\equiv \frac{\kappa a_0}{\pi G \Sigma_0} = \frac{2}{\alpha}(x_{\rm t0}+x_{\rm A0})^{1/2},\\
\beta_0 &\equiv x_{\rm t0}/x_{\rm A0},
\end{align}
where we denote $a_0^2 = {\rm v}_{\rm t0}^2 + {\rm v}_{\rm A0}^2$, as the average square value of effective sound speed.

\section{TASS State}\label{SecBasic}
The TASS state of the problem consists of a large scale spiral shock which has been investigated extensively in the literature. The solution depends on the five galactic background parameters (Group I). In this section, we investigate the dependence of the basic state on these parameters and present some observational applications. We first focus on the effects of gas self-gravity and magnetic field. Next, we show how the streaming motions and the time delay of star formation can tell us about the basic state of the problem. In the following discussion, except for the comparison to the self-gravitating solution in \citetalias{1986ApJ...309..496L}, a reference model for M$\,$51 is adopted with the rotation curve in \citet{1999ApJ...523..136S} and the pattern speed $\Omega_{\rm p} \simeq 40\,{\rm km\,s^{-1}\,kpc^{-1}}$ from \citet{2004ApJ...607..285Z}. The parameters and properties of the reference model are listed in Table \ref{c3table3}.

\begin{table}[!htb]
\begin{center}
\caption[Parameters in the Reference Model]{Parameters in the Reference Model}\label{c3table3}
\begin{tabular}{cc}
\hline
$\varpi$ & 2.0$\,$kpc \\
$L_{\rm arm}$ & 2.25$\,$kpc \\
$i_{\rm pitch}$ & $21.0^{\rm o}$ \\
$\tilde{L}$ & 2.61 \\
$\Omega_{\rm p}$ & 40.0$\,{\rm km\,s^{-1}\,kpc^{-1}}$ \\
$\Omega$ & 127$\,{\rm km\,s^{-1}\,kpc^{-1}}$ \\
$\kappa$ & 186$\,{\rm km\,s^{-1}\,kpc^{-1}}$ \\
$\rm v_{t0}$ & 10.0$\,{\rm km\,s^{-1}}$ \\
$B_{\varphi0}$ & 10.0$\,{\rm \mu G}$ \\
$F$ & $\sim 8\%$ \\ % 7.7
\hline
$\nu$ & -0.933 \\
$f$ & 0.2 \\
$x_{\rm t0}$ & 0.022 \\
$x_{\rm A0}$ & 0.02 \\
\hline
\end{tabular}
\end{center}
\end{table}

\subsection{Self-gravity of the Gas}
The gas dynamics is greatly affected by the self-gravity when the gas surface density $\Sigma_0$ is high. Theoretically, in order to obtain a self-consistent solution to the gas response calculation together with the Poisson equation for gas self-gravity, an iteration technique is required \citepalias{LS2012}. The presence of the shock and the sonic point in a steady-state calculation suggests that any spectral method requires some special care, which otherwise would not handle the shock jump and sonic point correctly. On the other hand, because of the long-range nature of gravity, the gravitation potential at each location depends on the gas density at all locations (which suggests the use of Fourier transformation). Therefore, to successfully obtain a steady-state solution, we start from the non-self-gravitating case and gradually increase $\alpha$ to integrate a new solution after we obtained the gravitational force from the previous solution. Convergence of the solution can be obtained until the value of $\alpha$ reaches some maximum value. We suspect the absence of time-steady solution for large $\alpha$ is related to the chaos through overlapping of the resonances similar to what was found in \citetalias{SMR1973} for strong spiral forcing \citep[c.f.,][]{2003ApJ...596..220C,2004csg..book.....S}.

The high density of gas can also give back-reaction to the stellar spiral density wave. Theoretically, both stars and gas should be treated equally in solving the Poisson equation \citep[e.g.,][which leads to nonlinear stellar density wave]{1971ApJ...166...37V}. Using linear WKB theory of the stellar waves (\citealt{1969ApJ...155..721L}; \citetalias{1986ApJ...309..496L}), one can simplify the relation between stars and gas. By neglecting the back-reaction to the stars we can set the strength of the stellar spiral forcing ($f$) and the self-gravity of the gas ($\alpha$) independently \citepalias{1986ApJ...309..496L}. In the following, the TASS profile is calculated using the galactic parameters in the solar neighborhood, and is compared to the model C in \citetalias{1986ApJ...309..496L}, where the gas surface density is $10\%$ of the stellar surface density at the spiral arm.

\begin{figure}[!htb]
\begin{center}
\includegraphics[scale=0.45]{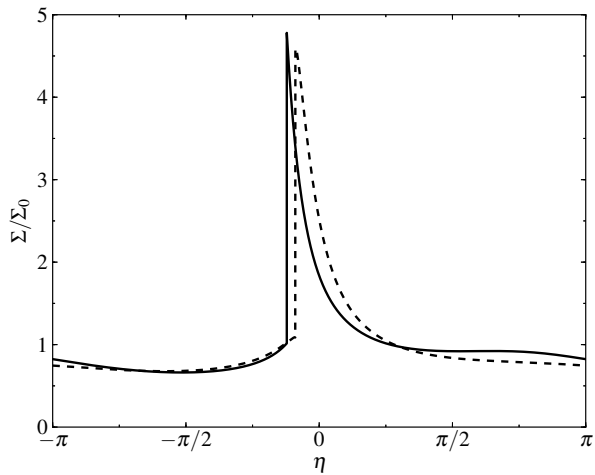} %c3fig1.eps} %fig5b.eps
\caption[Normalized TASS Profiles with the Presence of Gas Self-gravity]{Normalized TASS profiles with the presence of gas self-gravity ($\alpha=0.13$, dashed line) and the case without ($\alpha=0$, solid line). The mean surface density is $\Sigma_0$ so that the area under each curve is 1. The horizontal axis is the displacement from the minimum location of the stellar spiral potential.
}\label{c3fig1}
\end{center}
\end{figure}

The same set of galactic parameters of model C in \citetalias{1986ApJ...309..496L} is used with the exception of the parameters $x_{\rm t0}$, $x_{\rm A0}$, and $\alpha$. Because of the adoption of logatropic EOS, the turbulent gas pressure in the compression region is lower than the corresponding thermal pressure using the typical value of $8\,{\rm to}\,10$ $\rm km\,s^{-1}$ of turbulent (or thermal sound) speed. This leads to higher compression at the shock in our calculation ($4\,{\rm to}\,5$ times compared to the $2\,{\rm to}\,3$ times of the mean surface density in their paper). The maximum converged values of $\alpha$ in our calculations are lower than the corresponding $\alpha=0.26$ in their paper (based on the $10\%$ gas-star mass ratio in the spiral arm). Note that the limit on $\alpha$ also depends on the strength of stellar spiral potential $f$, in which a stronger stellar spiral potential can support a larger amount of gas for steady-state solution. Qualitatively, we find that the increase in $\alpha$ leads to a spiral shock at a further downstream location, but does not always increase the peak amount of relative compression ($\Sigma_{\rm peak}/\Sigma_0$) in the spiral arm. In particular, $\Sigma_{\rm peak}/\Sigma_0$ decreases in the self-gravitating case when the gas is weakly magnetized (e.g., $\beta_0=20$ in Figure \ref{c3fig1}). This agrees with the non-magnetic calculations in \citetalias{1986ApJ...309..496L} that show a decrease of relative peak compression for increasing $\alpha$. However, in terms of dimensional unit (${\rm M_\odot\,pc^{-2}}$), the peak surface density is generally increasing with $\alpha$. On the other hand, viscosity and back-reaction on the stellar spiral potential by the gas have been shown to smoothen the shock and lead to weaker compression. 

\subsection{Magnetic Field}
The magnetic fields in the TASS state are assumed to be parallel to the streamlines due to a steady-state solution of the induction equation \citep{1970ApJ...161..887R} in the ideal MHD regime, where $\mathbf{B} \propto \Sigma \mathbf{u}$. This is a good representation of the regular magnetic field in the large scale \citep{2011MNRAS.412.2396F}, which are found to be aligned with the spiral arms. In this configuration, the magnetic field provides extra pressure against the spiral shock compression in the $\eta$-direction. Under the tight-winding assumption of the spiral arms, the magnetic tension which is proportional to $\sin{i}$ is ignored here, and hence the streamlines are closed (c.f., Equation (52) of \citetalias{LS2012}) in the local model. In general, the magnetic field help broaden the gaseous spiral arm and lower the shock strength in the TASS state. In the following subsections, we continue with the parameters for a simple model of M$\,$51 introduced in Section \ref{SecPhyScale}.

\begin{figure}[!htb]
\begin{center}
\includegraphics[scale=0.45]{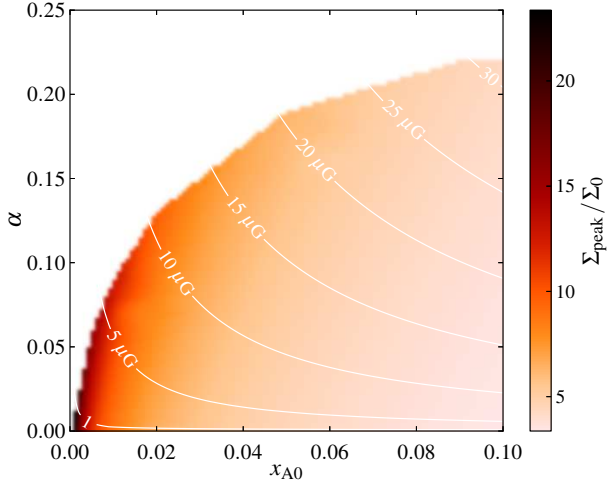}
\caption[Dependence of Normalized Gas Peak Surface Density on $x_{\rm A0}$ and $\alpha$]{The color-coded region shows the normalized peak surface density of the gas (i.e., maximum value of $1+\hat{\sigma}$). The horizontal and vertical axes are the magnetic and self-gravity parameters, respectively. The white contours represent the mean circular magnetic field ($\mu \rm G$). }\label{fig1a}
\end{center}
\end{figure}

\begin{figure}[!htb]
\begin{center}
\includegraphics[scale=0.45]{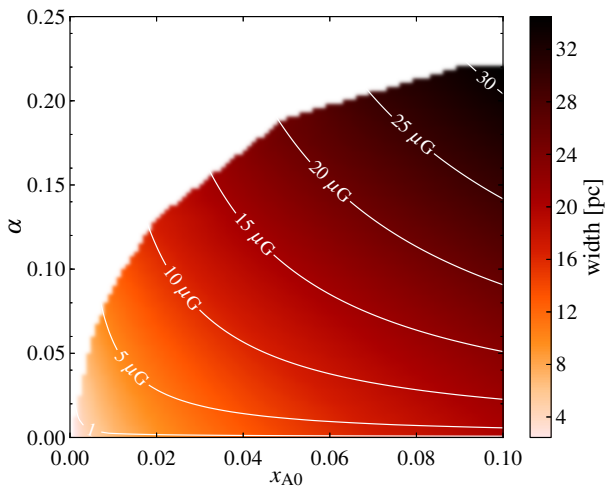}
\caption[Dependence of Spiral Arm Width on $x_{\rm A0}$ and $\alpha$]{Width of the gaseous spiral arm, defined as the distance from the shock location to the sonic point, at the same parameter space as Figure \ref{fig1a}.}\label{fig1b}
\end{center}
\end{figure}

\begin{figure}[!htb]
\begin{center}
\includegraphics[scale=0.45]{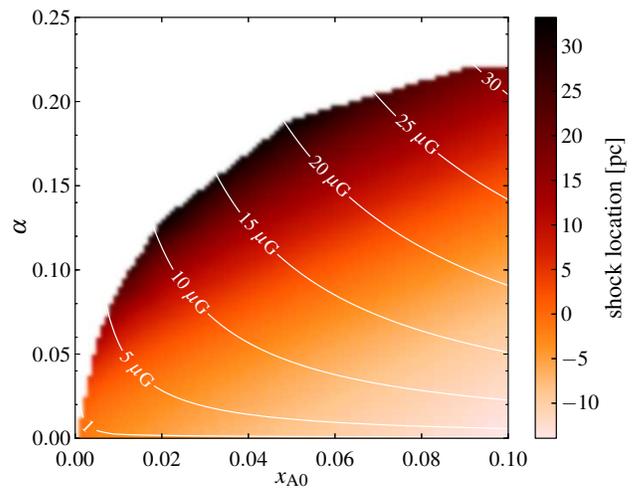}
\caption[Dependence of Shock Location on $x_{\rm A0}$ and $\alpha$]{Shock location ($\eta_{\rm sh}$) as the distance of the minimum location of the stellar spiral potential ($\eta=0$) at the same parameter space as Figure \ref{fig1a}. The arm-to-arm distance $L_{\rm arm} = 2.25\,{\rm kpc}$.}\label{fig1c}
\end{center}
\end{figure}

Figures \ref{fig1a}, \ref{fig1b}, and \ref{fig1c} show the effects of the parameters $x_{\rm A0}$ (magnetic) and $\alpha$ (self-gravity) on the normalized peak surface density of the gas, the width of the spiral arm, and the shock location of the TASS states, respectively. We overlay them with the white contours showing the strength of average magnetic field $B_{\varphi0}$, which is proportional to $(\alpha x_{\rm A0})^{1/2}$ in Equation (\ref{eq:defB0}). The blank region on the top-left corner has no steady-state solution for the combination of the parameters. \citet{KO2002} reported that the spiral shock in their simulations is oscillating about the stellar spiral potential in this high gas density regime. The color gradient of each figure is used to show which parameter is more important in determining the quantity of the color bar.

In addition, the upper boundary of the color region is interpreted as the line of maximum allowed values of $\alpha$ along the Alfv\'en's speed parameter $x_{\rm A0}$. Such sub-linear behavior of the upper limit $\alpha_{\rm max}$ suggests that the local gravitational instability in the spiral arm prohibits the existence of steady spiral shock with strong self-gravity. In other words, the magnetic field has a stabilizing effect against the self-gravity by providing extra pressure. In general, the relative gas density of the shock is more sensitive to the amount of magnetic pressure than self-gravity. 

As $x_{\rm A0}$ is lowered, the normalized peak surface density increases (Figure \ref{fig1a}) and the shock gets narrower (Figure \ref{fig1b}). There is a lower limit of $x_{\rm A0}$ for steady-state solution. This is partly because of our ``soft" equation of state cannot provide enough pressure at high density in the non-magnetic case. The purpose of logatropic EOS is to mimic the observed velocity dispersion that is lower at high gas density, and provide an uniform effective sound speed over a range of density with the consideration of magnetic field \citepalias{LS2012}. On the other hand, if an isothermal EOS is used, steady-state solutions exist without magnetic field as in previous studies.

We define the gaseous arm width using the perpendicular distance between the locations of the shock and the sonic point, which reads
\begin{align}
W = (\eta_{\rm mp} - \eta_{\rm sh}) L_0, 
\end{align}
where $\eta_{\rm mp}$ and $\eta_{\rm sh}$ are the $\eta$-coordinate of the magneto-sonic point and the shock, respectively. In Figure \ref{fig1b}, the magnetic field strength correlates well with the width of the gaseous spiral arm near the upper boundary (i.e., $B \propto W$ in this regime). The shock location $\eta_{\rm sh}L_0$ (Figure \ref{fig1c}) has a stronger dependence on $\alpha$ than $x_{\rm A0}$, in which higher value of $\alpha$ trends to shift the shock {\it downstream}. Note that the actual location of the spiral shock relative to the stellar spiral arm depends on other parameters as well (e.g., $\nu$). Such shifting due to gas self-gravity suggests that the prediction from purely hydrodynamic calculation \citep{2004MNRAS.349..909G} may systematically overestimate the corotation radius (or underestimate $\Omega_{\rm p}$).

\begin{figure}[!htb]
\begin{center}
\includegraphics[scale=0.45]{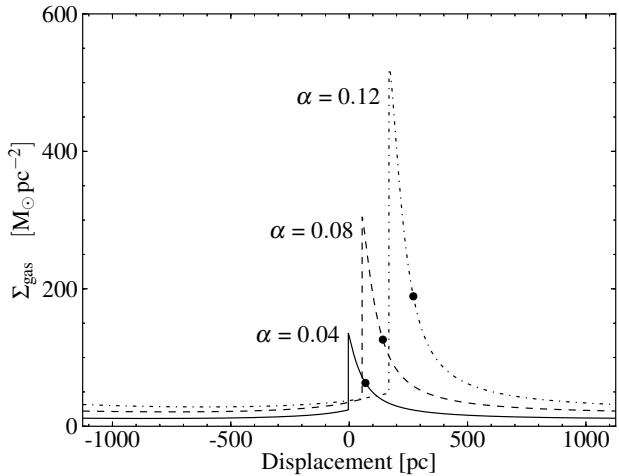}
\caption[Typical TASS Profiles for Different Values of $\alpha$]{Typical TASS profiles for different values of $\alpha$ (at $x_{\rm A0}=0.02$). These curves based the parameters which lie on a vertical line on the parameter space in the Figure \ref{fig1a}. The horizontal axis is the perpendicular displacement from the minimum location of the stellar spiral potential. The vertical axis is the dimensional gas surface density with the mean value at $\Sigma_0 = \alpha\Sigma_{\rm A}$. The black dots are the sonic point values.}\label{figm1_a}
\end{center}
\end{figure}

\begin{figure}[!htb]
\begin{center}
\includegraphics[scale=0.45]{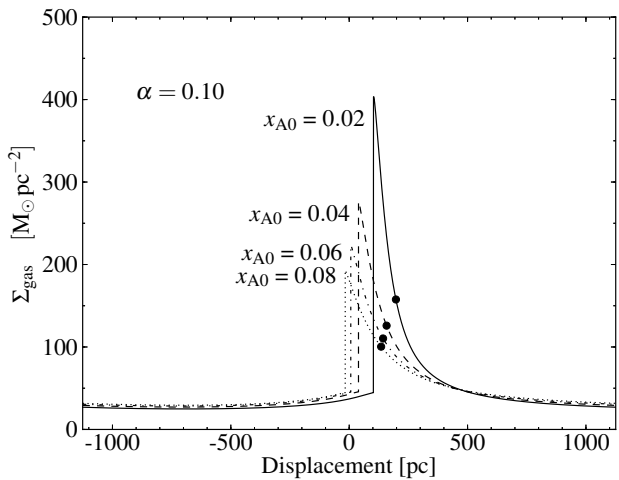}
\caption[Typical TASS Profiles for Different Values of $x_{\rm A0}$]{Typical TASS profiles for different values of $x_{\rm A0}$, with the black dots indicating the sonic points.}\label{figm1_xa}
\end{center}
\end{figure}

The bottom-left part of the parameter space is the most relevant when we adopt the realistic values of the magnetic field (${\sim}10 \, {\rm \mu G}$) and gas surface density (${\sim} 50\,{\rm M_\odot\, pc^{-2}}$). In this particular model with a sharp spiral shock, the width of the gaseous spiral arm is $10\%$ or less of $L_{\rm arm}=2.25\,{\rm kpc}$. Using $(x_{\rm A0}, \alpha)=(0.02, 0.10)$ as reference parameters, the shock profiles with variation in $\alpha$ and $x_{\rm A0}$ are shown in Figure \ref{figm1_a} and \ref{figm1_xa}, respectively. These two cases represent the solutions along a vertical line and a horizontal line on the $x_{\rm A0}$-$\alpha$ space.

In Figure \ref{figm1_a}, the area under each curve is proportional to the total gas mass between the spiral arms, and thus it is proportional to $\alpha$. As we indicated previously, the shock location is shifted downstream with stronger self-gravity. Similarly, Figure \ref{figm1_xa} shows that the stronger magnetic field leads to a weaker, wider, and more upstream spiral shock. 

\subsection{Streaming Motion}\label{subsec:streaming}

\begin{figure}[!htb]
\begin{center}
\includegraphics[scale=0.45]{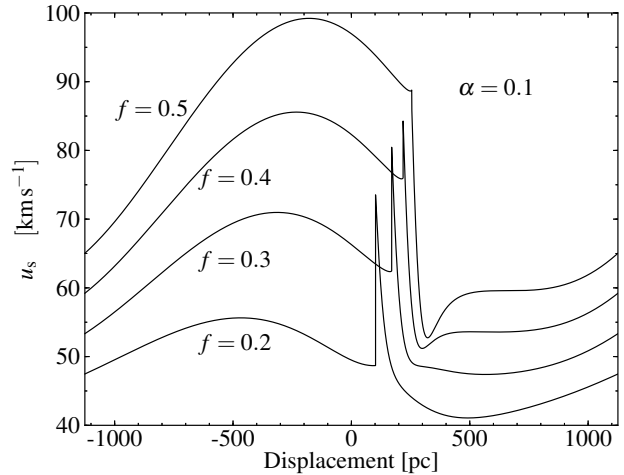} %us_f_p40.eps} % us_f.eps
\caption[Typical Profiles for the Magnitude of Streaming Motion]{Typical profiles for the magnitude of streaming motion along the perpendicular displacement from the spiral arm.}\label{fig2}
\end{center}
\end{figure}

The streaming motion is characterized by the non-circular motion of the gas due to the gravitational forcing of the spiral arm. In particular, the post-shock gas velocity indicates how strong the spiral shock and the corresponding shear near the spiral arm. The magnitude of such streaming velocity can be defined as the difference between the fluid velocity of the TASS state and the background circular velocity:
\begin{align}
u_{\rm s} = | \mathbf{u} - \mathbf{u}_0| = \sqrt{u_{\eta1}^2 + u_{\xi1}^2},
\end{align}
where $\mathbf{u}_0 = \varpi(\Omega-\Omega_{\rm p})\hat{e}_\varphi$ is the circular velocity in the pattern frame; $u_{\eta1}$ and $u_{\xi1}$ are the nonlinear perturbation due to spiral potential in the $\eta$- and $\xi$-directions, respectively. We show the profiles of $u_{\rm s}$ for the dimensionless strength of spiral forcing $f$ between 0.2 and 0.5 ($F=8-20\%$) in Figure \ref{fig2}. The narrow, sharp peaks correspond to the spiral shock (discontinuity in $u_{\eta1}$) in this particular model with low effective sound speed ($x_{\rm t0}=0.022$ and $x_{\rm A0}=0.02$). Except in the region near the shock, the magnitude of the streaming velocity varies gradually before or after the gas passing through the spiral shock. The magnitude of the streaming velocity scales roughly with $f$. As the streaming velocity can be obtained from observations \citep[e.g.,][]{1999ApJ...522..165A,2007ApJ...665.1138S, 2013ApJ...779...45M}, this provides an independent estimate of the spiral forcing which can be used to compare with the arm-interarm contrast due to the spiral structure from the stellar mass map data \citep{2011ApJ...737...32E}. For example, the model with $f=0.2$ gives a typical streaming velocity of $50\,\rm km\,s^{-1}$.

\subsection{Arm-Crossing Time}\label{subsec:arm}
In the quasi-static spiral structure (QSSS) picture, a newly-formed star drifts from its formation side in the shock because of velocity difference between the gas flow and the spiral density wave \citep{Roberts1969a}. Such displacement between the young stars (traced by H-$\alpha$ emission) and gas (e.g., HI or $\rm H_2$) is often referred as ``geometrical" offset \citep{2006A&A...458..441P,2008AJ....136.2872T,2009ApJ...697.1870E,2011ApJ...735..101F,2013ApJ...763...94L}, because of the assumption that the gas flow is almost circular (i.e., $\Delta \phi \propto \Delta t$). However, such assumption may underestimate of the age of young stars $t_{\rm SF}$ or overestimate $\Omega_{\rm p}$ \citep{2009ApJ...707.1650M} by ignoring the non-circular motion. The estimated migration time for a cloud or star to move from the shock to its current position depends sensitively on the strength of the spiral shock. Moreover, the age of the young star may be underestimated in the circular model by a factor of 4. The calculation of the migration time can be estimated using the velocity of the TASS flow.

The physical time-scale for migration from one spiral arm to the next arm is set by the total arm-crossing time $t_{\rm cross} = 2\pi/m(\Omega-\Omega_{\rm p})$, which is the same as in the absence of a spiral perturbation. It is a constant among different parameters because the reciprocal of the dimensionless $\eta$-velocity $1/u_\eta \propto \Sigma/\Sigma_0$ is a periodic function due to the closure of streamlines. Therefore, the spiral perturbation of the TASS state does not change the total arm-crossing time. By using the TASS state solution, the time for a cloud to move from the spiral shock to a location $\eta$ is given by
\begin{align}\nonumber
t(\eta) &= \int^{\eta}_{\eta_{\rm sh}} \frac{L_0}{u_\eta(\eta)}d\eta \\\label{time1}
&= \frac{1}{m(\Omega-\Omega_{\rm p})} \int^{\eta}_{\eta_{\rm sh}}\left(\frac{\Sigma}{\Sigma_0}\right)d\eta,
\end{align}
where $\eta_{\rm sh}$ is shock location. The integral can be further evaluated using the TASS state solution, which gives $(\eta-\eta_{\rm sh}) + [\hat{{\rm v}}(\eta) - \hat{{\rm v}}(\eta_{\rm sh})]$, where $\hat{{\rm v}}=u_{\xi1}/\sqrt{2UV}$ is the $\xi$-component of velocity due to the spiral perturbation. The integrand of Equation (\ref{time1}) implies that the stronger the shock, the longer the time for the fluid to reach the interarm region. On the other hand, the super-magnetosonic flow in the pre-shock region will compensate the time spent getting out the spiral arm, and keep the total time constant.

\begin{figure}[!htb]
\begin{center}
\includegraphics[scale=0.4]{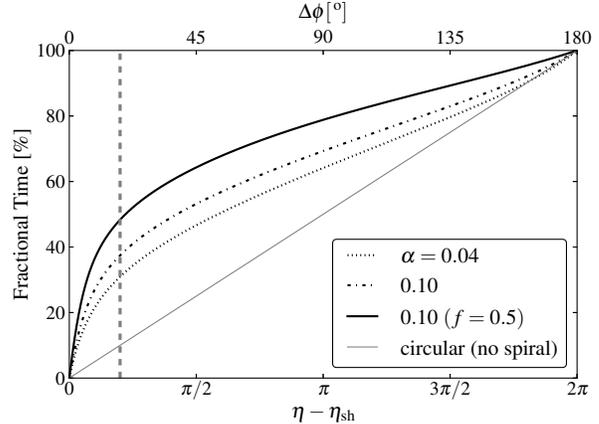} %fig30c_p40f2.eps} %c3s44fig1.eps
\caption[Fractional Arm-crossing Time for a Cloud]{Fractional arm-crossing times for a cloud moving from the shock to the interarm region. The vertical thick gray dashed line locates the position at $10\%$ arm-to-arm distance away from the shock. The total crossing time is $t_{\rm cross} = 35.3\,{\rm  Myr}$.}\label{figarm1} % omega_p = 40
\end{center}
\end{figure}

In Figure \ref{figarm1}, we show the fractional arm-crossing time $t(\eta)/t_{\rm cross}$ for a cloud moving from the shock to the interarm region using the aforementioned reference model. For the reference model (Table \ref{c3table3}), the post-shock flow is very slow ($u_\eta\lesssim 10\,{\rm km\,s^{-1}}$) in the spiral arm with a strong shock. In particular, it takes about half of the crossing time to travel only 10\% of the arm-to-arm distance (vertical dashed line). The time due to the circular flow is shown as the gray diagonal across the figure. On the top horizontal axis, the angular offset $\Delta \phi$, which is the amount of rotation (w.r.t. the galactic center) needed for matching the two patterns, is shown. The presence of a spiral shock reduces the flow speed and increases the time for displacement near the shock by a factor of 3-4. In other words, for a fixed value of star-formation time $t_{\rm SF}$, the amount of displacement of a cloud is {\it smaller} with a spiral shock and this would lead to smaller offset. Therefore, while the offsets of different tracers can be directly measured across the spiral arm, the flow time between two positions depends on the actual model adopted (e.g., circular motions, or spiral shock scenario).
 
In summary, the expected offsets among the location of minimum spiral potential $\eta_{\rm min}$, the location of spiral shock $\eta_{\rm sh}$, and the location of young stars $\eta_{\rm SF}$ are reduced by two effects: 1) the gas self-gravity makes the spiral shock more downstream and closer to $\eta_{\rm min}$; 2) the streaming motion reduces the perpendicular distance traveled by the young stars. As a result, the detailed modeling presented here may be useful to interpret the small offsets found in observation \citep[e.g.,][]{2006A&A...458..441P,2013ApJ...763...94L}.

\subsection{Summary of Parameter Study of TASS states}

We briefly summarized this section into the following points: 1) The self-gravity pushes the spiral shock downstream and enhances the (absolute) peak gas surface density. 2) The magnetic field decreases the strength of the spiral shock significantly by providing extra pressure. 3) The streaming motion can be used to characterize the strength of the spiral structure. 4) The streaming motion extends the duration of the gas flowing through the spiral arm significantly. 

\section{Feathering Instability}\label{SecFeather}

We perform the stability analysis of the feathering perturbation on top of a TASS state. A parameter study is presented with the focus on the effects of the gas self-gravity and magnetic field. The method of solution is previously presented in \citetalias{LS2012}. Here we only briefly describe the theoretical procedures.  Readers who are more interested in the results may skip to Section \ref{secGR}.

In the linear analysis, we study the perturbation for a single (positive) $\xi$-wavenumber $l$, such that it varies as ${\rm exp}\,i(\omega \tau - l\xi/\tilde{L})$, where $\omega$ is the complex frequency and $\tilde{L}=\cot{i}$. A set of linear ordinary differential equations (ODEs) can be obtained. Each solution has four complex Fourier components, each being a function of $\eta$: $\tilde{\sigma}_l(\eta)$, $\tilde{u}_l(\eta)$, $\tilde{{\rm v}}_l(\eta)$, and $\tilde{A}_{l}(\eta)$, where $\tilde{\sigma}_l$, $\tilde{u}_l$, and $\tilde{{\rm v}}_l$ have their usual meanings and $\tilde{A}_{l}$ is the $z$-component of the perturbational magnetic vector potential. We denote the quasi-one-dimensional TASS state and perturbational variables with hat and tilde, respectively. Such complex treatment allows each component to have a different phase difference (in $\xi$) with respect to each other. 

Assuming the system is periodic along the spiral arm ($\xi$-direction), the spacing of feathers (i.e., separation between peaks of density enhancement) resulted from a perturbation with a single wavenumber $l$ is given by
\begin{align}\label{eq:larm}
\lambda_{\rm feather} = \left(\frac{\tilde{L}}{|l|}\right)L_{\rm arm},
\end{align}
where $l/\tilde{L}$ is the effective wavenumber. Note that the wavenumber $l$ takes on an integer value, and asymptotically equals to the number of feathers to be found on a spiral arm from 0 to 180 degrees for a two-arm spiral structure (i.e., $360/m$ degrees for $m$-arm spirals). We define $\omega_{\rm T} \equiv \omega - (l/\tilde{L})\hat{{\rm v}}_{\rm T}$ to be the dimensionless Doppler-shifted frequency in the moving frame of the TASS flow along the spiral arm (c.f., Section 4 of \citetalias{LS2012}). In practice, we take $\hat{{\rm v}}_{\rm T}=-\nu/\tan{i}=u_{\xi0}/\sqrt{2UV}$ as we previously assumed for the TASS flow. Thus, $\omega_{\rm T}$ is an unknown complex eigenvalue to be determined by solving the ordinary differential equations (ODEs) and imposing the perturbational shock jump conditions as boundary conditions. 
We adopt the same reference model presented in the last section, which is based on the galactic parameters of the inner part of M\,51 (i.e., $\varpi=2\,{\rm kpc}$). The parameters of both turbulent gas and Alfv\'en's speed are relatively small (i.e., 0.02). The self-gravity parameter is set to $\alpha=0.1$ (i.e., $\Sigma_0 = 46\,{\rm M_\odot\,pc^{-2}}$).
%The spiral forcing parameter is $f=0.2$. 
%The radial dependence of the galactic parameters is discussed in Appendix \ref{appendix3A}.

\subsection{General Properties}

The solution of the feathering perturbations is computed by using a Fortran solver for boundary value problems (BVP\_SOLVER) described in \citet{shampine2006user} and \citet{Boisvert:2013:RBS:2427023.2427028}. We include the detail of transform the equations into standard form of boundary value problems in Appendix \ref{appendixAA}. For each value of $l$, we calculate the eigenfunctions and the corresponding complex eigenvalues, $\omega_{\rm T}$. There are generally multiple eigenvalues and eigenfunctions for each wavenumber. In particular, there are multiple branches of solution (e.g., instead of being a single ``continuous" function of $l$, the complex frequency $\omega_{\rm T}$ sometimes bifurcates or forms cusp) due to the existence of different waves in the MHD system (e.g. acoustic, Alfv\'en's wave, etc). To ensure that we follow the solution of the same branch, we increase $l$ with non-integral steps of increment (e.g., 0.01). Here we study the branch of solution with unstable modes at the immediate values of $l/\tilde{L}$ (around 3 to 5, which corresponds to a few hundred pc of the feather spacing as suggested by observations).

\begin{figure}
\begin{center}
\includegraphics[scale=0.45]{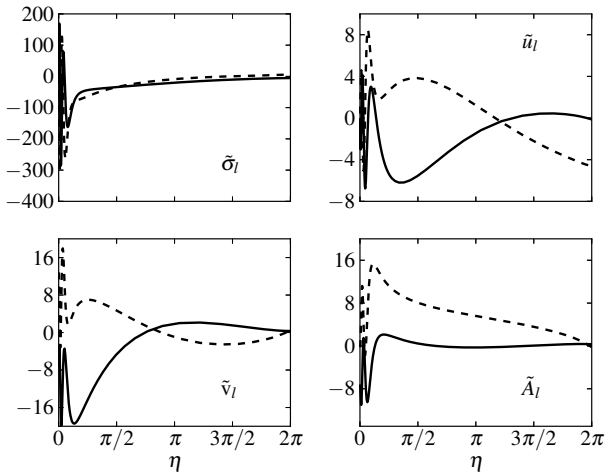} %fig51b-sola1-l11.nc.eps} 
\caption[Unstable Mode at $l/\tilde{L}=4.22$]{Unstable mode at $l/\tilde{L}=4.223$ (or $l=11$). The eigenfunctions (from top to bottom, from left to right) are $\tilde{\sigma}_l$, $\tilde{u}_l$, $\tilde{{\rm v}}_l$, and $\tilde{A}_{l}$ ($z$-component of the magnetic vector potential). The solid and dashed lines represent the real and the imaginary parts, respectively. The $\eta$-axis is measured from the shock front and the amplitude is chosen such that $\tilde{\sigma}(0) = 1$.
}\label{fig:eigenf2l11}
\end{center}
\end{figure}

An example of complex eigenfunctions of the perturbation is presented in Figure \ref{fig:eigenf2l11}. This is the fastest growing unstable (positive growth rate) mode for $f=0.2$ at $l/\tilde{L}=4.22$ (or $l=11$). The arbitrary complex multiplicative constant of the linear perturbation is chosen such that $\tilde{\sigma}_l(\eta)$ is $1+0i$ immediately after the shock (i.e., zero of the $\eta$-axis). In general, the solution varies rapidly in the beginning and decreases slowly for larger $\eta$. Also, the end points of the solution are not necessarily zero because of the shock jump conditions. 

\subsection{Growth Rates}\label{secGR}
\begin{figure}[!htb]
\begin{center}
\includegraphics[scale=0.45]{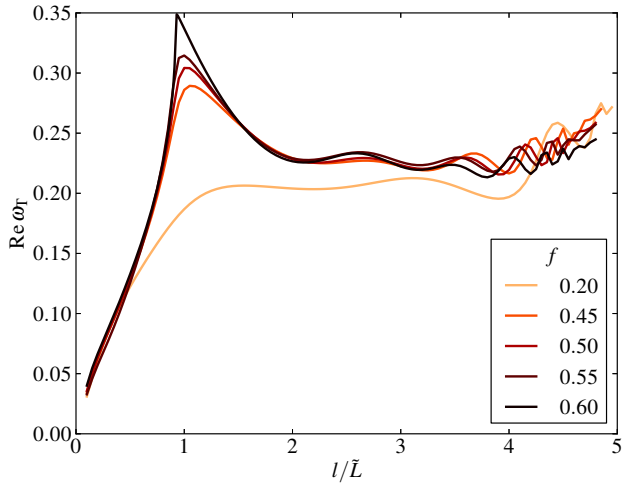} %omr_f.eps}
\caption[Real Part of $\omega_{\rm T}$ along $l$]{Real part of $\omega_{\rm T}$ along $l$ for some values of $f$. At $f=0.6$, the discontinuity at $l/\tilde{L}\simeq 1$ corresponds to a degeneracy between two branches (other branch not shown, see Figure \ref{fig:f6}). Darker lines for larger $f$.}\label{fig:omr_f}
\end{center}
\end{figure}

\begin{figure}[!htb]
\begin{center}
\includegraphics[scale=0.45]{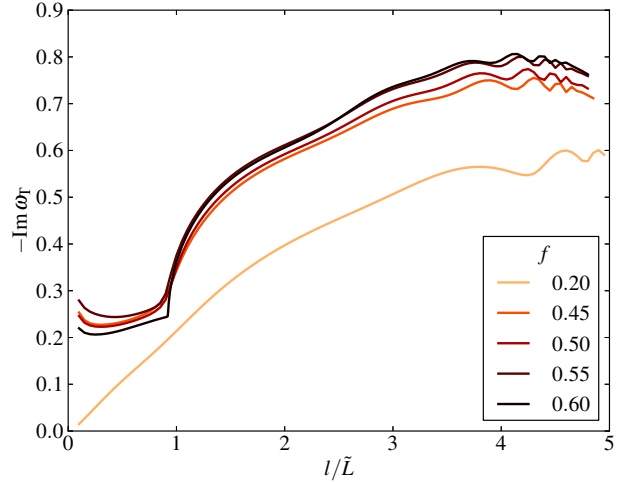} %omi_f.eps}
\caption[Growth Rate along $l$]{Growth rate (-Im($\omega_{\rm T}$)) along $l$ with same parameters as Figure \ref{fig:omr_f}. Darker lines for larger $f$.}\label{fig:omi_f}
\end{center}
\end{figure}

We study the stability of the feathering perturbation by exploring numerically the behavior of the perturbation frequency ($\omega_{\rm T}$) versus the effective wavenumber ($l/\tilde{L}$). There is no analytical dispersion relation because we make no approximation on the relative length scale in the $\eta$-direction between the nonlinear TASS state and the perturbation. As seen in the eigenfunctions of the perturbation (Figure \ref{fig:eigenf2l11}), we find no rapid oscillation along the $\eta$-direction (except very near to the shock for large $l$ cases), which suggests the integration in $\eta$ is necessary. In Figures \ref{fig:omr_f} and \ref{fig:omi_f}, we show respectively, the real part of frequency and the growth rate for a range of spiral forcing $f$. 

\begin{figure}[!htb]
\begin{center}
\includegraphics[scale=0.4,trim=1in 1in 1in 0in]{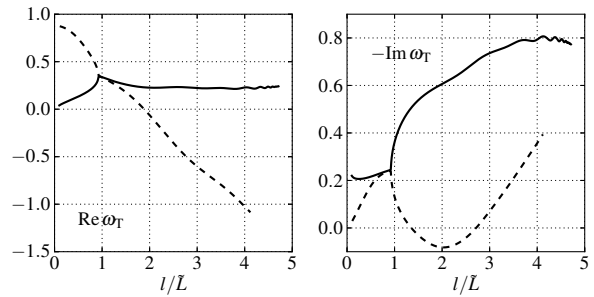} %f6freq.eps}
\caption{Real frequency ({\it left}) and growth rate ({\it right}) for $f=0.6$. The solid and dashed lines are two different branches of solution.}
\label{fig:f6}
\end{center}
\end{figure}

As inferred by the streaming motion estimation in the Section \ref{subsec:streaming}, we choose $f=0.2$ (or $F=8\%$) as a reference. At this level of spiral forcing, only one branch of mode is found. Both real and imaginary parts of $\omega_{\rm T}$ are close to zero when $l \to 0$, which suggests its TASS state is non-oscillating and stable to the plane-parallel perturbation ($l=0$). However, we should only interpret the low $l$ regime with caution as the WKBJ approximation of the solution of Poisson equation breaks down (Equation (\ref{eq:poissonsol})). At higher values of $f$, we find at least two branches of frequency, but we show only the branch which is unstable for moderate  wavenumber here (and has similar behavior as $f=0.2$). At $f=0.6$, two branches cross at $l/\tilde{L}=0.93$ (see Figure \ref{fig:f6}) and it appears as a cusp in Figure \ref{fig:omr_f}. One of the branches reaches the stable regime (${\rm Im}\,\omega_{\rm T} > 0$), while other branch remains unstable and reaches a maximum growth rate. The cusp at crossing of the two branches may be related to the phenomenon of ``avoided crossing", which also appears in a similar system of a massive protoplanetary disk \citep{2014arXiv1404.1923L}. If that is the case, some characters of the branches may exchange. However, we will leave this for future investigation. For moderate values of $l/\tilde{L}$, the real frequency is almost flat for the unstable branch. This corresponds to a family of perturbation that has a small group velocity relative to the local circular flow (i.e., $d\,{\rm Re}(\omega_{\rm T})/dl \simeq 0$). This particular feature of feathers was also found by \citet{KO2002} where the feathers move along the spiral arm in their numerical simulations.

The fastest growing mode is located at $l/\tilde{L} = 4.22$ where $l=11$. This number is not sensitive to the value of spiral forcing. This mode corresponds to a spacing of $530\,{\rm pc}$ between the feathers. The neighboring modes ($\Delta l=\pm 1$) give a $30\,{\rm pc}$ difference to this number. This feather separation agrees with the findings in \citet{2006ApJ...650..818L} (see Figure 21 in their paper). Also, the dimensionless growth rate of $\gamma = -{\rm Im}(\omega_{\rm T})= 0.807$ is similar to what we obtained in \citetalias{LS2012}. In general, the (dimensional) $e$-folding time $t_0$ of unit growth rate ($\kappa t_0=1$) is
\begin{align}
t_0 = \frac{1}{\kappa} = \frac{m(\Omega-\Omega_{\rm p})}{2\pi\kappa}t_{\rm cross} = \left(\frac{-\nu}{2\pi}\right)t_{\rm cross},
\end{align} 
where $t_{\rm cross}=2\pi/m(\Omega-\Omega_{\rm p})$ is the arm-crossing time in the pattern frame. The number of $e$-folds of growth per unit arm-crossing time is $N=\gamma(2\pi/-\nu)$, where $-\nu$ is defined in Equation (\ref{eq5}). At $\varpi=2\,{\rm kpc}$ where $-\nu=0.933$ and $t_0 = 0.148\,t_{\rm cross}$, the perturbation could grow by a factor of  $\exp{\gamma(t_{\rm cross}/t_0}) = e^{5.43} = 229$ for $\gamma = 0.807$ in one arm-crossing time. Combining the fact that the gas flows slowly inside the spiral arm (Section \ref{subsec:arm}), this rapid growth of feathering provides a favorable condition for star formation \citep{2014ApJ...780...32E}.

\begin{figure}[!htb]
\begin{center}
\includegraphics[scale=0.4]{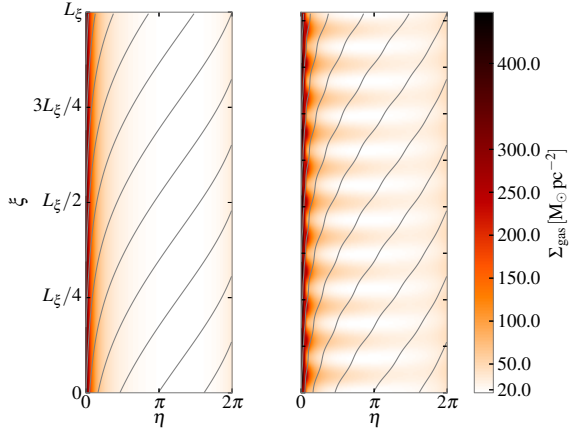} %f2a1l11-2D.eps}
\caption[Plot of the Surface Density of the Background Flow and the Perturbed Flow of the Most Unstable Mode]
{Plot of the surface density of the background flow (\textit{left}) and the perturbed flow of the most unstable mode (\textit{right}, $l=11$). The arbitrary amplitude of the perturbation is 0.01 in this case. The scale of the vertical axis is $L_\xi = L_{\rm arm}/\tan{i}$, where $i=21^{\rm o}$. The grey contours are the magnetic field lines.}
\label{fig:f2a1l11-2D}
\end{center}
\end{figure}

In Figure \ref{fig:f2a1l11-2D}, we show the two-dimensional gas surface density for the background and the one with feathering perturbation with an arbitrary amplitude $\epsilon=0.01$. In this case, the contrast between the gas surface density inside the feathers and inter-feathers is around 4 to 6, with a lower value farther away from the spiral arm. The dimensional $\eta$-velocity fluctuation (along $\xi$) due to the feathers is given by
\begin{align}
\Delta u_\eta = \epsilon |\tilde{u_l}| \sqrt{2UV},
\end{align}
where $|\tilde{u_l}|$ is the magnitude of the $\eta$-velocity perturbation, $\sqrt{2UV}$ is the dimensional velocity scale in $\eta$-direction. At $\varpi=2\,{\rm kpc}$ where $\sqrt{2UV}=66.7\,{\rm km\,s^{-1}}$, the $\eta$-velocity fluctuation is $5\,{\rm km\,s^{-1}}$. On the other hand, we find that the $\xi$-velocity fluctuation is of the same order of magnitude.

\subsection{Dependence on Self-gravity}

\begin{figure}[!htb]
\begin{center}
\includegraphics[scale=0.45]{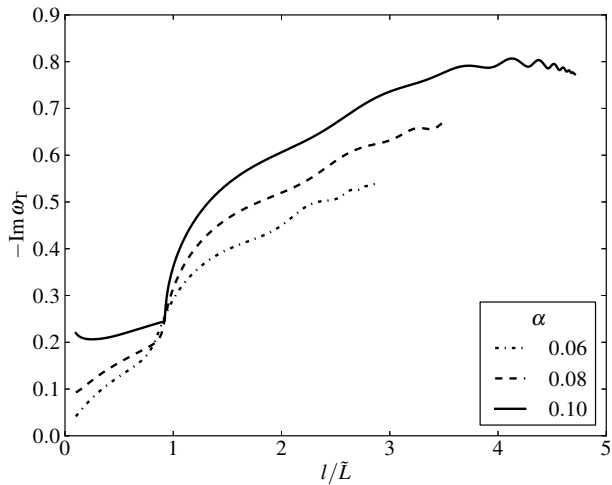} %omi-alpha.eps}
\caption[Growth Rate for different $\alpha$]{Growth rate for different values of $\alpha$ for $f=0.6$ and $x_{\rm A0}=0.02$. }\label{fig:omi_alpha}
\end{center}
\end{figure}

The dependence of the gas self-gravity on the feather instability is similar to that of the spiral forcing. We find that the stronger the self-gravity, the higher the growth rate. In Figure \ref{fig:omi_alpha}, we show the growth rate of the reference model with $f=0.6$ at $\alpha=0.06$, 0.08, and 0.1. The value $f$ is tuned up to allow a wider range of $\alpha$ of the TASS states. Comparing this to Figure \ref{fig:omi_f} (varying $f$), the cusp at $l/\tilde{L}=0.93$ becomes a smooth transition when $\alpha$ is lowered. Also, the solutions become numerically unstable at a larger value of $l/\tilde{L}$. This numerical artifact is partly due to the increase in stiffness of the ODEs of the perturbation. We expect it is solvable by considering a proper matching condition at the critical point of the ODEs \citep[see,][]{2014ApJ...789...68K}, and leave this for future investigation. In any case, the increase in growth rate with self-gravity is expected for a perturbation caused by the gravitational instability.

\subsection{Dependence on Magnetic Field}\label{subsec:magnetic}

\begin{figure}[!htb]
\begin{center}
\includegraphics[scale=0.45]{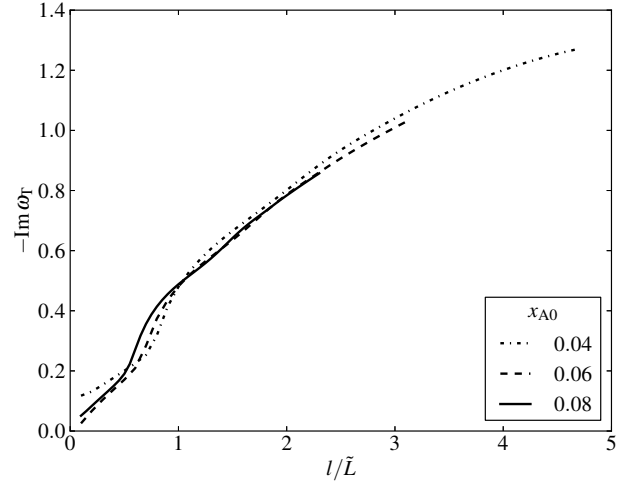} %omi-xa-f6.eps}
\caption[Growth Rate for different $x_{\rm A0}$]{Growth rate for different values of $x_{\rm A0}$ at $f=0.6$ and $\alpha=0.2$. }\label{fig:omi_xa}
\end{center}
\end{figure}

We study the dependence of the magnetic field by varying the dimensionless parameter $x_{\rm A0}$. In general, stronger the magnetic field, the weaker and wider the spiral shock. However, despite the changes in the TASS state, the growth rates have the very similar behavior (Figure \ref{fig:omi_xa}) for the range of $x_{\rm A0}$ explored (i.e., 0.04 to 0.08, corresponding to the mean plasma beta $\beta_0=2\,{\rm to}\,4$). As the increase of $x_{\rm A0}$ leads to a weaker shock, we study the solutions at the higher value of $\alpha=0.2$ (instead of 0.1 in the reference model) in order to maintain a moderate value of the shock strength. Although the growth rate does not show a maximum in this set of parameters, the curves in Figure \ref{fig:omi_xa} still provide an estimate for the lower limit of the most unstable wavenumber, which is $l/\tilde{L}\simeq 5$. This corresponds to $\lambda_{\rm feather}\simeq 450\,{\rm pc}$ in linear scale. Therefore, combining the finding that $x_{\rm A0}$ does not change the growth rate within the range we explored, we can conclude that the wavelength of the fastest growing mode decreases with stronger self-gravity. 

\begin{figure}[!htb]
\begin{center}
\includegraphics[scale=0.45]{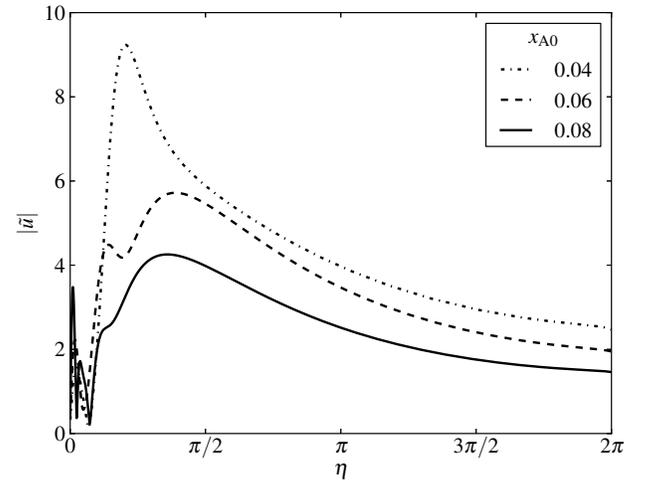} %ut-xa-f6.eps}
\caption[$|\tilde{u}|$ for different $x_{\rm A0}$]{Magnitude of $\eta$-velocity perturbation ($|\tilde{u}|$) for different values of $x_{\rm A0}$ at $l/\tilde{L}=2$, $f=0.6$ and $\alpha=0.2$. }\label{fig:ut_xa}
\end{center}
\end{figure}

Note that the real part of the frequency $\omega_{\rm T}$ and the eigenfunctions do change with $x_{\rm A0}$ accordingly. As an example, the magnitudes of the perturbational $\eta$-velocity at $l/\tilde{L}=2$ is shown in Figure \ref{fig:ut_xa}. The fluctuation decreases with the increasing strength of magnetic field. As these modes have very similar growth rate, we are comparing the strength of perturbation at same time instant assuming their initial amplitudes are the same.

\subsection{Dependence on Radius}
\begin{figure}[!htb]
\begin{center}
\includegraphics[scale=0.45]{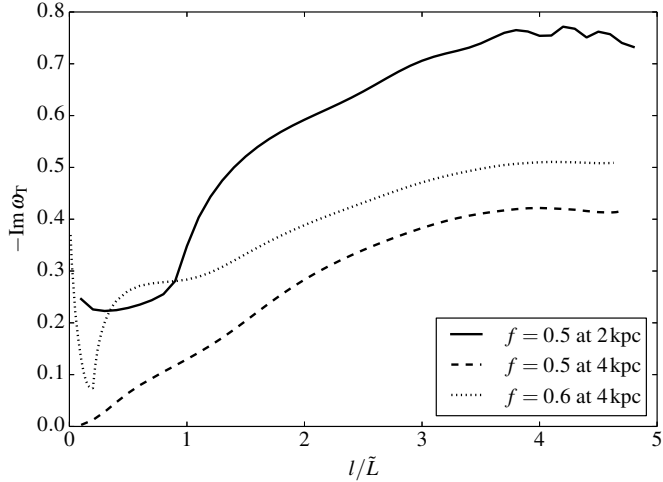} %ut-xa-f6.eps}
\caption[Growth Rate for Different $\varpi$]{Growth rate for different values of spiral forcing at $\varpi=2$ and $4\,{\rm kpc}$.}\label{fig:omi_radius}
\end{center}
\end{figure}

The local calculation is performed at a different radius to provide insights on how the substructure forms at different locations. Here we compare the growth rate at $\varpi=4\,{\rm kpc}$ to the reference model ($2\,{\rm kpc}$) in Figure \ref{fig:omi_radius}. For easier comparison of theoretical models, we only change the radius and spiral forcing while keeping the same self-gravity and Alfv\'en speed parameters. At a larger radius, the value of Doppler-shifted frequency $-\nu$ is smaller. The TASS state is more likely to have a secondary density enhancement when the location is near to the ultra-harmonic resonance \citepalias{SMR1973}. The feathering instability has a lower growth rate and peaks at a slightly smaller effective wavenumber ($l/\tilde{L}=3.84$). Using Equation (\ref{eq:larm}), the feather spacing is $1\,{\rm kpc}$, which qualitatively agrees with the finding that spacing increases with radius in \citet{2006ApJ...650..818L}. The use of more realistic values of parameters requires adaptation of a global model of M\,51, such as a radial profile of magnetic field strength. Therefore, computing a radial trend of feather spacing is more meaningful when making comparison to global simulations in the future.

\section{Discussion}\label{SecDiscuss}

\subsection{Summary of Results}\label{SecSummary}

The dependence of the TASS state and the feathering instability on the various parameters is investigated. There are five parameters for the TASS state (Group I) and two additional parameters (Group II) for the feather instability. We are most interested in the effects of the magnetic field and the self-gravity (c.f., Section \ref{SecBasic}), because the Group II parameters such as $\Omega/\kappa$ and pitch angle of the spiral arm can be obtained confidently from observations, while the effects of the sprial forcing $f$ and the relevant frequency of the spiral structure $\nu$ were studied theoretically in the literature \citepalias[e.g.,][]{SMR1973, 1986ApJ...309..496L}. In the first part, we studied a few observationally relevant quantities and we summarize these with their most important determining factor(s) in Table \ref{tab:summary}.

\begin{threeparttable}[!htb]
\begin{center}
\caption[Determining Factors of Various Quantities]{Determining factors of various quantities in the TASS state}\label{tab:summary}
\footnotesize
\begin{tabular}{cc}
\hline
relative peak surface density ($\Sigma_{\rm g, peak}/\Sigma_0$) & spiral forcing (f), \\
& pressure \tnote{a} \\
absolute gas surface density ($\Sigma_{\rm g}$) & self-gravity ($\alpha$) \\
spiral arm width (W) & magnetic field \\
spiral shock location ($\eta_{\rm sh}$) & self-gravity ($\alpha$) \\
streaming velocity ($u_{\rm s}$) & spiral forcing (f) \\
\hline
\end{tabular}
\begin{tablenotes}
\item{(a)} both magnetic ($x_{\rm A0}$) and turbulent ($x_{\rm t0}$) gas pressure
\end{tablenotes}
\end{center}
\end{threeparttable}
\\

The stability analysis of the feathering perturbation shows that there exist growing unstable modes. Some of these modes look like feathers jutting out from the spiral arm (Figure \ref{fig:f2a1l11-2D}). Because of the nonlinear TASS state and the perturbed boundary conditions, we do not have an analytical dispersion relation (which is often obtained by neglecting boundary conditions and using WKB analysis). Instead, the complex frequency $\omega_{\rm T}$ is numerically computed for a range of effective wavenumber along the arm. A few branches of perturbations are found. The branch with large unstable growth rate has a group velocity close to (but not exactly equal to) the local rotation velocity. This indicates the feathers may also move along the spiral arm \citep{KO2002}. The most unstable mode is located at a wavelength of around $530\,{\rm pc}$ in our M\,51 model, which agrees with the spacing of feathers in observations, such as \citet{2006ApJ...650..818L}. We also examine the parameter dependence of the feathering instability. Each parameter changes the underlying TASS state and thus the exact detail of the eigenfunctions of the perturbation. However, apart from the crossing phenomenon of branches and the increase of the growth rate (see Figure \ref{fig:omi_f}), the increase in spiral forcing $f$ alone does not vary the wavelength of the most unstable mode. On the other hand, the growth rate increases more sensitively with the self-gravity. For the magnetic field, the growth rates remain similar for a range of Alfv\'en speed parameter. We also examine the instability at a larger radius, in which the growth rate is lowered. Our calculation suggest the feather separations increase with radius, unless the fastest growing wavenumber change significantly.

\paragraph{Feather Spacing and Jeans Length}
% Jeans Length is the stability critieria, not the most unstable mode. The most unstable wavenumber = 1/2 k_J
The local Jeans length is commonly used to compare with the feather or spur spacing in the numerical simulations and observations. The two-dimensional local Jeans length inside the spiral arm, which is given by
\begin{align}
\lambda_{\rm J} = \frac{a_0^2}{G\Sigma_{\rm spiral}} = \left(\frac{x_0}{\alpha}\right)\left(\frac{\Sigma_0}{\Sigma_{\rm peak}}\right)L_{\rm arm},
\end{align}
where $\Sigma_{\rm spiral}=\Sigma_{\rm peak}$ is the peak surface density of the spiral arm, and $x_0 = x_{\rm t0} + x_{\rm A0}$ is the dimensionless counterpart of the square of mean effective sound speed $a_0^2$ (c.f., Section \ref{SecPhyScale}).  Typical ratio between $\lambda_{\rm feather}$ and $\lambda_{\rm J}$ is less than 10, where a larger ratio is obtained for weaker self-gravity when considering the vertical stratification \citep{2006ApJ...646..213K}. In our calculation of the fastest growing mode in Section \ref{secGR}, the ratio is 
\begin{align}
\frac{\lambda_{\rm feather}}{\lambda_{\rm J}} = \left(\frac{\tilde{L}}{l}\right) \left(\frac{\alpha}{x_0}\right)\left(\frac{\Sigma_{\rm peak}}{\Sigma_0}\right)  \simeq 7.9,
\end{align}
where $l/\tilde{L}=4.22$ and $\Sigma_{\rm peak}/\Sigma_0\simeq13$ for the density compression of the spiral shock. While this ratio lies within the range found in \citet{KO2002}, \citet{2006ApJ...650..818L} showed a large scatter for this ratio, partly because of the uncertainty in deriving the Jeans length from the gas surface density. As $\Sigma_{\rm peak}/\Sigma_0$ is sensitive to the magnetic pressure as well (Section \ref{SecBasic}), we expect this ratio may also have some radial variation that is different among galaxies. Therefore, we hope our calculation of the wavelength of the fastest growing mode can be used as a better diagnostic in future measurements.

\subsection{Applications}
\paragraph{Comparison to Other Instabilities}

In order to explain the substructure in spiral arms, other mechanisms such as wiggle instability \citep[e.g.,][]{2004MNRAS.349..270W} were proposed. The stability analysis of a corrugated spiral shock for the purely hydrodynamical case without self-gravity nor magnetic field by \citet{2014ApJ...789...68K} showed that such wiggle instability is related to the generation of potential vorticity at the deformed shock front and that the small scale perturbation grows fastest (e.g., 7\% of arm-arm distance in their example case). While this may be the case in some galaxies (in particular, for non-regular substructure), our analysis shows self-gravitating feathering instability can also occur without the background shear due to the differentially rotating galactic disk. Our assumption differs from the previous analytical studies using shearing coordinates \citep[e.g.,][]{1985ApJ...297...61B,1987ApJ...312..626E,B88,1994ApJ...433...39E}. The main reason not to include the galactic shear (and hence shearing-box boundary conditions) explicitly in our formulation is that a simple periodic boundary condition allows normal-mode analysis with the perturbation amplitude $\epsilon(t)$ instead of linear time $t$ as in a shearing-box. In addition, the shearing-periodic boundary condition along the $\xi$-axes is only asymptotically correct when the shearing-box is tilted against the circular direction (i.e., $\Omega=\Omega(\varpi)$ is not a constant along the spiral arm, see Figure \ref{fig:coord}). On the other hand, the shearing-box approximation differs from our formulation only in the background velocity along the spiral arm. Therefore, we expect our results are qualitatively the same if we were to adopt such approximation. In any case, to make progress from the local approximation, global numerical simulation of a galaxy is probably a better tool to properly compare the features among the feathering instability and other shearing instabilities. Careful investigation is needed to understand and quantify the difference of the self-gravitating MHD instability in \citet{2006ApJ...647..997S} and the shearing instability found in purely hydrodynamical simulations \citep[e.g.,][]{2004MNRAS.349..270W, 2006MNRAS.367..873D, 2006ApJ...647..997S, 2014MNRAS.440..208K}.

On the other hand, the analysis suggests that in some cases, the primary shock is rippled, with little density variation, rather than producing high density contrast feathers. As all single-mode perturbations vary sinusoidally along the spiral arm, such kind of perturbation has a concentrated fluctuation near the shock (see, e.g., Figure \ref{fig:eigenf2l11}) and has a significantly lower amplitude away from the spiral arm. In some cases, low $l$-mode (with large wavelength) may correspond to \citet{1979ApJ...231..372E} analysis on dust lanes collapsing along its length, which give rise to kpc-scale separation between massive cloud complexes or the ``beads on a string" phenomenon.

Furthermore, the difference between the observed curved structure of feathers and the straight single-mode density fluctuation seen in Figure \ref{fig:f2a1l11-2D} may be due to the nonlinear mode-coupling at late times, which is similar to the formation of mushroom structure in the Rayleigh-Taylor instability. We speculate that the bending of feathers may be due to the difference in growth rate and pitch angle to the spiral arm of each $l$-mode. Individual $l$-mode grows in amplitude separately until some nonlinear quasi-static equilibrium is reached. Such scenario may occur in two stages: The large $l$-mode perturbation develops first and dominates near the spiral shock. When the perturbations at the downstream flow start to develop, the small $l$-modes with smaller pitch angle to the spiral arm will mix with the large $l$-modes and result in bending feature. Therefore, in the current picture of feathering instability, the bending structure is not kinematic shearing of material features, but the mode coupling of unstable waves that adopts as its basis of expansion the summation of different modal quantum numbers in a linearized description. This hypothesis can be readily tested in the future by comparing local nonlinear ``shear-less" simulations and shearing-box simulations such as \citet{KO2002} and \citet{2014ApJ...789...68K}.

\section{Conclusion}\label{Sec:Conclusion}

This paper provides some theoretical understanding of the feathering phenomenon near the spiral arms. Complemented by the results from previous simulations and a recent paper on the purely-hydrodynamic case \citep{2014ApJ...789...68K}, the feathering instability with self-gravity and magnetic field is likely to be the formation mechanism of feathers. The magnetic field, which is approximately parallel to the spiral arm, suppresses the wiggle instability while provides a preferential direction for the gravitational collapse along the field lines. On the other hand, previous simulations suggest that feathers do not form without including self-gravity of the gas. 

\paragraph{Comparison to Observations}\
\citet{2006ApJ...650..818L} summarized a list of characteristics of feathers from their archival study of optical images. In general, the normal modes of feathering instability in this paper, such as the one in Figure \ref{fig:f2a1l11-2D}, match the general feather characteristics found in their paper (c.f., conclusion section), namely, 1) feathers extend from the spiral shock (or primary dust lines) to the interarm regions with large pitch angles; 2) feathers are often associated with density clumps (or bright star-forming regions), especially near their beginnings in the spiral arm; 3) feathers coalesce and extend to the next arm. There are other characteristics that are not reproduced in our analysis because of the local approximation, namely, pitch angle and curvature of the feathers. Some interesting morphologies such as beads-on-a-string may be explained by the long-wavelength mode which are not covered in this study. On the other hand, the lattice structure of feathers (i.e., appearance of both perpendicular and parallel dust lanes) may depend on the base state. For example, second density enhancement parallel to the main spiral arm can form near the ultra-harmonic resonance \citepalias{SMR1973} or under some conditions of the rotation curve \citep{2014MNRAS.440..208K}. In a boarder context, the instability associated with shocks may explain the cases even if the base state is not described by the TASS framework, such as flocculent galaxies \citep[e.g.,][]{2012ApJ...757..155R} and barred galaxies \citep[e.g.,][]{2006ApJ...650..818L}.

\paragraph{Remarks for Future Observations and Tests}
Our analysis suggests that the feather instability is sensitive to the base state of the spiral arms under the TASS picture. As a result, this may be related to the detection characteristics found in \citet{2006ApJ...650..818L}, such as feathers are most common in Sb-Sc type galaxies but not other spiral types. Galaxies with prominent primary dust lanes but without any feathers may indicate low gas surface density with strong magnetic field, as suggested by the simulations in \citet{2006ApJ...647..997S}. Measurement of feather spacing of more real galaxies and its relation to the Jeans length remain crucial for testing different theories and numerical simulations. In particular, the spacing of the feathers measured along the spiral arm reflect underlying variations of self-gravity and magnetic field (Section \ref{SecFeather}). The radial variation of spiral forcing $F$, which is often assumed a constant, should also be considered in the calculation \citep{2014ApJ...785..103F}. Combining the information obtained from the measurements of quantities in Table \ref{tab:summary}, a better understanding can be gained on the pattern speed ($\Omega_{\rm p}$) and magnetic field which are more difficult to measure generally.

\acknowledgements
This research is part of the author's PhD Thesis Dissertation in the Physics Department of UCSD. WKL thanks Frank Shu for his guidance over the graduate study. WKL also acknowledges Art Wolfe at UCSD who recently passed away for insightful discussions. WKL is grateful for the suggestions and comments from Ron Allen, Bruce Elmegreen, Woong-Tae Kim, Ron Taam, and Hsiang-Hsu Wang. The author acknowledges the support of the Theoretical Institute for Advanced Research in Astrophysics (TIARA) based in Academia Sinica's Institute of Astronomy and Astrophysics (ASIAA). This research has made use of the NASA/IPAC Extragalactic Database (NED), which is operated by the Jet Propulsion Laboratory, California Institute of Technology, under contract with the National Aeronautics and Space Administration.

\appendix

\section{Equations for Standard Boundary Value Problem Solver}\label{appendixAA}
In this appendix, we provide the governing equations for the feathering instability to facilitate the use of numerical solvers for the boundary value problem (BVP) that are available to the public \citep[e.g.,][]{Boisvert:2013:RBS:2427023.2427028}. The differential algebraic equations in the problem \citepalias[c.f.,][]{LS2012} are transformed into the standard form of BVP (Equations (\ref{eq:perteq}) and (\ref{eq:pertbc})). This appendix may be of interest to complement this work and \citetalias{LS2012} for actual calculation of the characteristic frequency and eigenfunctions of the feathering instability.

The governing equations of the perturbation can be written as a set of linear ODEs which has the following form:
\begin{align}\label{eq:perteq}
\mathbf{A}_l(\eta)\frac{d\bm{y}_l}{d\eta} = \mathbf{B}_{\omega_{\rm T}, l}(\eta)\bm{y}_l,
\end{align}
where $\mathbf{A}_l(\eta)$ and $\mathbf{B}_{\omega_{\rm T}, l}(\eta)$ are $4\times4\,$complex matrices depending on the TASS states (e.g., $\hat{u}(\eta)$), and $\bm{y}_l(\eta) = [\tilde{\sigma}_l, \tilde{u}_l, \tilde{{\rm v}}_l, \tilde{A}_{l}]^{\rm T}$ is a column vector of perturbational variables evaluated at $\eta$. The boundary conditions are given by a system of linear equations (the subscript $l$ is omitted for clarity):
\begin{align}\label{eq:pertbc}
\mathbf{Q}_1\bm{y}_{1} + \bm{\beta}_1 \epsilon = \mathbf{Q}_2\bm{y}_{2} + \bm{\beta}_2 \epsilon,
\end{align}
where $\mathbf{Q}_i$ and $\bm{\beta}_i$ are a $4\times4\,$matrix and a column vector depending on the TASS state, respectively. The subscript $i=1,2$ represents the pre-shock and post-shock locations. The arbitrary amplitude $\epsilon$ in this linear treatment is assumed to be one in the calculation, but is set to a small value when adding the perturbation to the TASS state. As discussed in the Appendix A of \citetalias{LS2012}, the solution of the perturbational Poisson equation (for self-gravity) is expressed as the Fourier-transformed dimensionless gravitational potential in the following form
\begin{align}\label{eq:poissonsol}
\tilde{\phi}_l(\eta) = -\frac{\tilde{\sigma_l}(\eta)}{|l/\tilde{L}|},
\end{align}
where the WKBJ approximation is applied. We look for the perturbations with a positive growth rate which grow exponentially and lead to nonlinear development of the overdense regions. The method of solution is discussed in detail in \citetalias{LS2012}. The major step to derive Equation (\ref{eq:perteq}) is to eliminate the second derivative of $\tilde{A}_l(\eta)$ in the momentum equations by the use of the perturbational induction equation. After some algebra and using the TASS state equations, we obtain the coefficients of Equation (\ref{eq:perteq}). The matrix $\mathbf{A}_l(\eta)$ is given by
\begin{align}\label{eq:appA}
\begin{pmatrix}
		u_{\rm T} & \sigma_{\rm T} & 0 & 0 \\ 
		\hat{b}_l & u_{\rm T} - x_{\rm A0}\frac{\sigma_{\rm T}}{{u_{\rm T}}} & 
		\frac{\kappa}{2\Omega}\tan{i}\frac{x_{A0}}{u_{\rm T}} & -x_{\rm A0}\shat^\prime/\sigma_{\rm T}  \\ 
		0 & \frac{2\Omega}{\kappa}x_{\rm A0}\frac{\tan{i}}{u_{\rm T}} & u_{\rm T}+x_{\rm A0}\tan^2 i/\nu & 0 \\
		0 & 0 & 0 & u_{\rm T} \\
	\end{pmatrix},
\end{align}
where $u_{\rm T} = -\nu+\hat{u}=u_{\eta}/\sqrt{2UV}$, $\sigma_{\rm T}=1+\hat{\sigma}=\Sigma/\Sigma_0$, $\hat{b}_l = x_{\rm t0}/\sigma_{\rm T}^2 - \alpha/|l/\tilde{L}|$, and the prime denotes the $\eta$-derivative of the TASS state. As defined previously, $x_{\rm t0}$ and $x_{\rm A0}$ are the square of normalized turbulent sound speed and Alfv\'en speed, respectively. The matrix $\mathbf{B}_{\omega_{\rm T}, l}$ is given by the following elements:
\begin{align}
B_{11} &= -\uhat^\prime-i\omega_{\rm T}, \\
B_{12} &= -\shat^\prime,\\
B_{13} &= i(l/\tilde{L})(\kappa/2\Omega)\sigma_{\rm T},\\
B_{14} &= 0,\\
B_{21} &= 2x_{\rm t0}\shat^\prime/\sigma_{\rm T}^3,\\
B_{22} &= -\uhat^\prime-i\omega_{\rm T} + \frac{x_{\rm A0}\sigma_{\rm T}}{u^2_{\rm T}}\left(-2\uhat'-i\omega_{\rm T}\right), \\
B_{23} &= 1 + x_{\rm A0}\frac{(\uhat'+i\omega_{\rm T})}{u^2_{\rm T}}\frac{\kappa}{2\Omega}\tan{i}, \\
B_{24} &= -x_{\rm A0}\left(\frac{l}{\tilde{L}}\right)^2
-\frac{x_{\rm A0}}{u_{\rm T}}\left[\left(\frac{il}{\tilde{L}}\right)\left(\frac{\kappa}{2\Omega}\right)\sigma_{\rm T}+
i\omega_{\rm T}\frac{(\uhat'+i\omega_{\rm T})}{u_{\rm T}}\right],\\
B_{31} &= (2\Omega/\kappa)(il/\tilde{L})\hat{b}_l,\\
B_{32} &= -\sigma_{\rm T} - \frac{2\Omega}{\kappa}\frac{x_{\rm A0}}{u^2_{\rm T}}\left(-2\uhat'-i\omega_{\rm T}\right)\tan{i}, \\
B_{33} &= -i\omega_{\rm T} -\frac{x_{\rm A0}}{-\nu}\frac{(\uhat'+i\omega_{\rm T})}{u_{\rm T}}\tan^2{i}, \\
B_{34} &= \left(\frac{2\Omega}{\kappa}\right)\frac{x_{\rm A0}}{\sigma_{\rm T}}\left[\tan{i}\left(\frac{l}{\tilde{L}}\right)^2-\shat'\frac{il}{\tilde{L}}\right] -\frac{2\Omega}{\kappa}\frac{x_{\rm A0}}{\sigma_{\rm T}}\left[\frac{il}{\tilde{L}}\left(\frac{\kappa}{2\Omega}\right)\sigma_{\rm T}+i\omega_{\rm T}\frac{(\uhat'+i\omega_{\rm T})}{u_{\rm T}}\right]\frac{\tan{i}}{u_{\rm T}} \\
B_{41} &= 0,\\
B_{42} &= \sigma_{\rm T},\\
B_{43} &= -(\kappa/2\Omega)\tan{i},\\
B_{44} &= -i\omega_{\rm T},
\end{align}
where $B_{ij}$ are the elements of $\mathbf{B}_{\omega_{\rm T}, l}$, and $\omega_{\rm T} = \omega - (l/\tilde{L})\hat{\rm v}_{\rm T}$. For the boundary conditions (perturbational shock jump conditions), a similar elimination procedure for $\tilde{A}''_l$ is required. Thus, the matrix $\mathbf{Q}_i$ in Equation (\ref{eq:pertbc}) is given by
\begin{align}
\begin{pmatrix}
		u_{\rm T} & \sigma_{\rm T} & 0 & 0 \\ 
		u_{\rm T}^2 + x_{\rm T0}/\sigma_{\rm T} & -2\nu- x_{\rm A0}\frac{\sigma_{\rm T}^2}{{u_{\rm T}}} & 
		\frac{\kappa}{2\Omega}\tan{i}x_{\rm A0}\frac{\sigma_{\rm T}}{u_{\rm T}} & i\omega_{\rm T}x_{\rm A0}\frac{\sigma_{\rm T}}{u_{\rm T}}  \\ 
		0 & x_{\rm A0}\tan{i}\frac{\sigma_{\rm T}}{{u_{\rm T}}} & -\frac{\kappa}{2\Omega}\left(\nu+x_{\rm A0}\tan^2{i}/u_{\rm T}\right) & x_{\rm A0}\left(\frac{il}{\tilde{L}}\sigma_{\rm T} + i\omega_{\rm T}\tan{i}/u_{\rm T}\right) \\
		0 & 0 & 0 & 1 \\
\end{pmatrix},
\end{align}
where $i=1,2$ denotes the evaluation at the each side of the shock. Lastly, the column vector $\bm{\beta}_i$ is given by
\begin{align}
\begin{pmatrix}
- i\sigma_{\rm T}\omega_{\rm T} \\
-\left[u_{\rm T}^2 - \frac{x_{\rm T0}}{\sigma_{\rm T}} - x_{\rm A0} \sigma_{\rm T}\right]\shat'+2i\nu\omega_{\rm T} \\
-x_{\rm A0}\tan{i}\shat' -\frac{\kappa}{2\Omega}\nu\shat + il\nu\uhat -ilx_{\rm A0}\sigma_{\rm T}^2 \\
\sigma_{\rm T}
\end{pmatrix}.
\end{align}

\section{Numerical Issues}\label{appendixBB}

Here we discuss some numerical issues in our calculation and how we possibly resolve them in future analysis. In the parameter study of the feathering instability, the effective wavenumber $l/\tilde{L}\lesssim 5$ is studied as the equations become less numerically stable for large $l$. One reason is that the stiffness of the Equation (\ref{eq:perteq}) increases with $l$ as the determinant of the ``mass-matrix" $\mathbf{A}_l$ (Equation (\ref{eq:appA})) is proportional (asymptotically) to:
\begin{align}
u_{\rm T}^2 - \frac{x_{\rm t0}}{1+\hat{\sigma}} - x_{\rm A0}(1+\hat{\sigma}) + \frac{\alpha}{|l/\tilde{L}|}(1+\hat{\sigma}),
\end{align}
where $u_{\rm T} = -\nu+\hat{u}=u_{\eta}/\sqrt{2UV}$ and $1+\hat{\sigma}=\Sigma/\Sigma_0$ are the dimensionless flow velocity perpendicular to the spiral arm in the pattern frame and the relative surface density, respectively. Except for the last term due to self-gravity, this quantity is related to the sonic point relation (i.e., $u_\eta^2-x=0$ where $x=a^2$ is the square of sound speed). Therefore, in front of the shock where the flow is sub-magnetosonic, the above determinant may be close to zero or even negative if the last term is not large enough (e.g., when the self-gravity parameter $\alpha$ is small or the effective wavenumber $l/\tilde{L}$ is large). However, the existence of such critical point (i.e., at certain $\eta$ where $\mathbf{A}_l$ is singular) is due to the assumption of a strictly $\xi$-periodic (and single-mode) flow and the WKBJ approximation of the Poisson equation in our analysis. We expect this regime (e.g., low gas surface density) can be studied with less analytical effort in numerical simulation, and we leave this for future investigation. Indeed, the recent paper by \citet{2014ApJ...789...68K} treated such critical point (or sonic point in their case) as an additional boundary condition. In any case, if we ambitiously perform WKB analysis on the equations (e.g., assuming a large $\eta$-wavenumber), we may derive a dispersion relation similar to the one for a differentially-rotating and self-gravitating disk. This means that the growth rate will eventually decline due to the stabilizing effect of the gas pressure in the small scale. In the mean time, we treat such numerical limit as the lower limit of the most unstable wavenumber.

Lastly, the approximation of razor-thin disk and WKBJ self-gravity overestimates the gravitational force \citep{2006ApJ...646..213K}. In principle, at small $l/\tilde{L}$ where WKBJ approximation breaks down, one can perform one more level of iterations to obtain a self-consistent solution of the Poisson equation (as an improvement over Equation (\ref{eq:poissonsol})). Since feathers are sub-kpc scale structure ($l/\tilde{L}>2$ by Equation \ref{eq:larm}), we do not worry such scenario. Thus, we limit our analytical interpretation on moderate values of $l/\tilde{L}$. 

\bibliography{p2}

\begin{thebibliography}{58}
\expandafter\ifx\csname natexlab\endcsname\relax\def\natexlab#1{#1}\fi

\bibitem[{{Aalto} {et~al.}(1999){Aalto}, {H{\"u}ttemeister}, {Scoville}, \&
  {Thaddeus}}]{1999ApJ...522..165A}
{Aalto}, S., {H{\"u}ttemeister}, S., {Scoville}, N.~Z., \& {Thaddeus}, P. 1999,
  \apj, 522, 165

\bibitem[{{Balbus}(1988)}]{B88}
{Balbus}, S.~A. 1988, \apj, 324, 60

\bibitem[{{Balbus} \& {Cowie}(1985)}]{1985ApJ...297...61B}
{Balbus}, S.~A., \& {Cowie}, L.~L. 1985, \apj, 297, 61

\bibitem[{{Bertin} \& {Lin}(1996)}]{1996ssgd.book.....B}
{Bertin}, G., \& {Lin}, C.~C. 1996, {Spiral structure in galaxies a density
  wave theory}

\bibitem[{Boisvert {et~al.}(2013)Boisvert, Muir, \&
  Spiteri}]{Boisvert:2013:RBS:2427023.2427028}
Boisvert, J.~J., Muir, P.~H., \& Spiteri, R.~J. 2013, ACM Trans. Math. Softw.,
  39, 11:1

\bibitem[{{Chakrabarti} {et~al.}(2003){Chakrabarti}, {Laughlin}, \&
  {Shu}}]{2003ApJ...596..220C}
{Chakrabarti}, S., {Laughlin}, G., \& {Shu}, F.~H. 2003, \apj, 596, 220

\bibitem[{{Corder} {et~al.}(2008){Corder}, {Sheth}, {Scoville}, {Koda},
  {Vogel}, \& {Ostriker}}]{2008ApJ...689..148C}
{Corder}, S., {Sheth}, K., {Scoville}, N.~Z., {Koda}, J., {Vogel}, S.~N., \&
  {Ostriker}, E. 2008, \apj, 689, 148

\bibitem[{{Dobbs} \& {Bonnell}(2006)}]{2006MNRAS.367..873D}
{Dobbs}, C.~L., \& {Bonnell}, I.~A. 2006, \mnras, 367, 873

\bibitem[{{D'Onghia} {et~al.}(2013){D'Onghia}, {Vogelsberger}, \&
  {Hernquist}}]{2013ApJ...766...34D}
{D'Onghia}, E., {Vogelsberger}, M., \& {Hernquist}, L. 2013, \apj, 766, 34

\bibitem[{{Egusa} {et~al.}(2009){Egusa}, {Kohno}, {Sofue}, {Nakanishi}, \&
  {Komugi}}]{2009ApJ...697.1870E}
{Egusa}, F., {Kohno}, K., {Sofue}, Y., {Nakanishi}, H., \& {Komugi}, S. 2009,
  \apj, 697, 1870

\bibitem[{{Elmegreen}(1979)}]{1979ApJ...231..372E}
{Elmegreen}, B.~G. 1979, \apj, 231, 372

\bibitem[{{Elmegreen}(1987)}]{1987ApJ...312..626E}
---. 1987, \apj, 312, 626

\bibitem[{{Elmegreen}(1994)}]{1994ApJ...433...39E}
---. 1994, \apj, 433, 39

\bibitem[{{Elmegreen}(1980)}]{1980ApJ...242..528E}
{Elmegreen}, D.~M. 1980, \apj, 242, 528

\bibitem[{{Elmegreen} {et~al.}(2011){Elmegreen}, {Elmegreen}, {Yau},
  {Athanassoula}, {Bosma}, {Buta}, {Helou}, {Ho}, {Gadotti}, {Knapen},
  {Laurikainen}, {Madore}, {Masters}, {Meidt}, {Men{\'e}ndez-Delmestre},
  {Regan}, {Salo}, {Sheth}, {Zaritsky}, {Aravena}, {Skibba}, {Hinz}, {Laine},
  {Gil de Paz}, {Mu{\~n}oz-Mateos}, {Seibert}, {Mizusawa}, {Kim}, \& {Erroz
  Ferrer}}]{2011ApJ...737...32E}
{Elmegreen}, D.~M., {et~al.} 2011, \apj, 737, 32

\bibitem[{{Elmegreen} {et~al.}(2014){Elmegreen}, {Elmegreen}, {Erroz-Ferrer},
  {Knapen}, {Teich}, {Popinchalk}, {Athanassoula}, {Bosma}, {Comer{\'o}n},
  {Efremov}, {Gadotti}, {Gil de Paz}, {Hinz}, {Ho}, {Holwerda}, {Kim}, {Laine},
  {Laurikainen}, {Men{\'e}ndez-Delmestre}, {Mizusawa}, {Mu{\~n}oz-Mateos},
  {Regan}, {Salo}, {Seibert}, \& {Sheth}}]{2014ApJ...780...32E}
---. 2014, \apj, 780, 32

\bibitem[{{Feng} {et~al.}(2014){Feng}, {Lin}, {Wang}, \&
  {Taam}}]{2014ApJ...785..103F}
{Feng}, C.-C., {Lin}, L.-H., {Wang}, H.-H., \& {Taam}, R.~E. 2014, \apj, 785,
  103

\bibitem[{{Fletcher} {et~al.}(2011){Fletcher}, {Beck}, {Shukurov},
  {Berkhuijsen}, \& {Horellou}}]{2011MNRAS.412.2396F}
{Fletcher}, A., {Beck}, R., {Shukurov}, A., {Berkhuijsen}, E.~M., \&
  {Horellou}, C. 2011, \mnras, 412, 2396

\bibitem[{{Foyle} {et~al.}(2011){Foyle}, {Rix}, {Dobbs}, {Leroy}, \&
  {Walter}}]{2011ApJ...735..101F}
{Foyle}, K., {Rix}, H.-W., {Dobbs}, C.~L., {Leroy}, A.~K., \& {Walter}, F.
  2011, \apj, 735, 101

\bibitem[{{Gittins} \& {Clarke}(2004)}]{2004MNRAS.349..909G}
{Gittins}, D.~M., \& {Clarke}, C.~J. 2004, \mnras, 349, 909

\bibitem[{{Julian} \& {Toomre}(1966)}]{1966ApJ...146..810J}
{Julian}, W.~H., \& {Toomre}, A. 1966, \apj, 146, 810

\bibitem[{{Kim} \& {Ostriker}(2002)}]{KO2002}
{Kim}, W., \& {Ostriker}, E.~C. 2002, \apj, 570, 132

\bibitem[{{Kim} {et~al.}(2014){Kim}, {Kim}, \& {Kim}}]{2014ApJ...789...68K}
{Kim}, W.-T., {Kim}, Y., \& {Kim}, J.-G. 2014, \apj, 789, 68

\bibitem[{{Kim} \& {Ostriker}(2006)}]{2006ApJ...646..213K}
{Kim}, W.-T., \& {Ostriker}, E.~C. 2006, \apj, 646, 213

\bibitem[{{Kim} \& {Kim}(2014)}]{2014MNRAS.440..208K}
{Kim}, Y., \& {Kim}, W.-T. 2014, \mnras, 440, 208

\bibitem[{{La Vigne} {et~al.}(2006){La Vigne}, {Vogel}, \&
  {Ostriker}}]{2006ApJ...650..818L}
{La Vigne}, M.~A., {Vogel}, S.~N., \& {Ostriker}, E.~C. 2006, \apj, 650, 818

\bibitem[{{Lee} \& {Shu}(2012)}]{LS2012}
{Lee}, W.-K., \& {Shu}, F.~H. 2012, \apj, 756, 45

\bibitem[{{Lin} \& {Bertin}(1995)}]{1995NYASA.773..125L}
{Lin}, C.~C., \& {Bertin}, G. 1995, Annals of the New York Academy of Sciences,
  773, 125

\bibitem[{{Lin} \& {Shu}(1964)}]{1964ApJ...140..646L}
{Lin}, C.~C., \& {Shu}, F.~H. 1964, \apj, 140, 646

\bibitem[{{Lin} {et~al.}(1969){Lin}, {Yuan}, \& {Shu}}]{1969ApJ...155..721L}
{Lin}, C.~C., {Yuan}, C., \& {Shu}, F.~H. 1969, \apj, 155, 721

\bibitem[{{Lin}(2014)}]{2014arXiv1404.1923L}
{Lin}, M.-K. 2014, ArXiv e-prints

\bibitem[{{Lizano} \& {Shu}(1989)}]{1989ApJ...342..834L}
{Lizano}, S., \& {Shu}, F.~H. 1989, \apj, 342, 834

\bibitem[{{Louie} {et~al.}(2013){Louie}, {Koda}, \&
  {Egusa}}]{2013ApJ...763...94L}
{Louie}, M., {Koda}, J., \& {Egusa}, F. 2013, \apj, 763, 94

\bibitem[{{Lubow} {et~al.}(1986){Lubow}, {Cowie}, \&
  {Balbus}}]{1986ApJ...309..496L}
{Lubow}, S.~H., {Cowie}, L.~L., \& {Balbus}, S.~A. 1986, \apj, 309, 496

\bibitem[{{Lynds}(1970)}]{1970IAUS...38...26L}
{Lynds}, B.~T. 1970, in IAU Symp. 38, The Spiral Structure of our Galaxy, ed.
  {W.~Becker \& G.~I.~Kontopoulos} (Cambridge: Cambridge Univ. Press), 26

\bibitem[{{Mart{\'{\i}}nez-Garc{\'{\i}}a}
  {et~al.}(2009){Mart{\'{\i}}nez-Garc{\'{\i}}a}, {Gonz{\'a}lez-L{\'o}pezlira},
  \& {G{\'o}mez}}]{2009ApJ...707.1650M}
{Mart{\'{\i}}nez-Garc{\'{\i}}a}, E.~E., {Gonz{\'a}lez-L{\'o}pezlira}, R.~A., \&
  {G{\'o}mez}, G.~C. 2009, \apj, 707, 1650

\bibitem[{{Meidt} {et~al.}(2013){Meidt}, {Schinnerer}, {Garc{\'{\i}}a-Burillo},
  {Hughes}, {Colombo}, {Pety}, {Dobbs}, {Schuster}, {Kramer}, {Leroy}, {Dumas},
  \& {Thompson}}]{2013ApJ...779...45M}
{Meidt}, S.~E., {et~al.} 2013, \apj, 779, 45

\bibitem[{{Patrikeev} {et~al.}(2006){Patrikeev}, {Fletcher}, {Stepanov},
  {Beck}, {Berkhuijsen}, {Frick}, \& {Horellou}}]{2006A&A...458..441P}
{Patrikeev}, I., {Fletcher}, A., {Stepanov}, R., {Beck}, R., {Berkhuijsen},
  E.~M., {Frick}, P., \& {Horellou}, C. 2006, \aap, 458, 441

\bibitem[{{Piontek} \& {Ostriker}(2005)}]{2005ApJ...629..849P}
{Piontek}, R.~A., \& {Ostriker}, E.~C. 2005, \apj, 629, 849

\bibitem[{{Rebolledo} {et~al.}(2012){Rebolledo}, {Wong}, {Leroy}, {Koda}, \&
  {Donovan Meyer}}]{2012ApJ...757..155R}
{Rebolledo}, D., {Wong}, T., {Leroy}, A., {Koda}, J., \& {Donovan Meyer}, J.
  2012, \apj, 757, 155

\bibitem[{{Roberts}(1969)}]{Roberts1969a}
{Roberts}, W.~W. 1969, \apj, 158, 123

\bibitem[{{Roberts} \& {Yuan}(1970)}]{1970ApJ...161..887R}
{Roberts}, Jr., W.~W., \& {Yuan}, C. 1970, \apj, 161, 887

\bibitem[{{Scoville} {et~al.}(2001){Scoville}, {Polletta}, {Ewald}, {Stolovy},
  {Thompson}, \& {Rieke}}]{2001AJ....122.3017S}
{Scoville}, N.~Z., {Polletta}, M., {Ewald}, S., {Stolovy}, S.~R., {Thompson},
  R., \& {Rieke}, M. 2001, \aj, 122, 3017

\bibitem[{{Sellwood}(2011)}]{2011MNRAS.410.1637S}
{Sellwood}, J.~A. 2011, \mnras, 410, 1637

\bibitem[{{Sellwood}(2012)}]{2012ApJ...751...44S}
---. 2012, \apj, 751, 44

\bibitem[{{Sellwood} \& {Carlberg}(2014)}]{2014ApJ...785..137S}
{Sellwood}, J.~A., \& {Carlberg}, R.~G. 2014, \apj, 785, 137

\bibitem[{Shampine {et~al.}(2006)Shampine, Muir, \& Xu}]{shampine2006user}
Shampine, L., Muir, P., \& Xu, H. 2006, JNAIAM, 1, 201

\bibitem[{{Shetty} \& {Ostriker}(2006)}]{2006ApJ...647..997S}
{Shetty}, R., \& {Ostriker}, E.~C. 2006, \apj, 647, 997

\bibitem[{{Shetty} {et~al.}(2007){Shetty}, {Vogel}, {Ostriker}, \&
  {Teuben}}]{2007ApJ...665.1138S}
{Shetty}, R., {Vogel}, S.~N., {Ostriker}, E.~C., \& {Teuben}, P.~J. 2007, \apj,
  665, 1138

\bibitem[{{Shu} {et~al.}(2004){Shu}, {Chakrabarti}, \&
  {Laughlin}}]{2004csg..book.....S}
{Shu}, F.~H., {Chakrabarti}, S., \& {Laughlin}, G. 2004, {Chaos in Spiral
  Galaxies} (Dordrecht: Kluwer Academic Publishers)

\bibitem[{{Shu} {et~al.}(1972){Shu}, {Milione}, {Gebel}, {Yuan}, {Goldsmith},
  \& {Roberts}}]{1972ApJ...173..557S}
{Shu}, F.~H., {Milione}, V., {Gebel}, W., {Yuan}, C., {Goldsmith}, D.~W., \&
  {Roberts}, W.~W. 1972, \apj, 173, 557

\bibitem[{Shu {et~al.}(1973)Shu, Milione, \& Roberts}]{SMR1973}
Shu, F.~H., Milione, V., \& Roberts, W.~W. 1973, \apj, 183, 819

\bibitem[{{Sofue} {et~al.}(1999){Sofue}, {Tutui}, {Honma}, {Tomita},
  {Takamiya}, {Koda}, \& {Takeda}}]{1999ApJ...523..136S}
{Sofue}, Y., {Tutui}, Y., {Honma}, M., {Tomita}, A., {Takamiya}, T., {Koda},
  J., \& {Takeda}, Y. 1999, \apj, 523, 136

\bibitem[{{Tamburro} {et~al.}(2008){Tamburro}, {Rix}, {Walter}, {Brinks}, {de
  Blok}, {Kennicutt}, \& {Mac Low}}]{2008AJ....136.2872T}
{Tamburro}, D., {Rix}, H.-W., {Walter}, F., {Brinks}, E., {de Blok}, W.~J.~G.,
  {Kennicutt}, R.~C., \& {Mac Low}, M.-M. 2008, \aj, 136, 2872

\bibitem[{{Vandervoort}(1971)}]{1971ApJ...166...37V}
{Vandervoort}, P.~O. 1971, \apj, 166, 37

\bibitem[{{Wada} \& {Koda}(2004)}]{2004MNRAS.349..270W}
{Wada}, K., \& {Koda}, J. 2004, \mnras, 349, 270

\bibitem[{{Yuan}(1969)}]{1969ApJ...158..889Y}
{Yuan}, C. 1969, \apj, 158, 889

\bibitem[{{Zimmer} {et~al.}(2004){Zimmer}, {Rand}, \&
  {McGraw}}]{2004ApJ...607..285Z}
{Zimmer}, P., {Rand}, R.~J., \& {McGraw}, J.~T. 2004, \apj, 607, 285

\end{thebibliography}

\end{document}